\documentclass[aps,prd,twocolumn,amsmath,amssymb,
  amsfonts,superscriptaddress,floatfix,nofootinbib,preprintnumbers]{revtex4-2}

\pdfoutput=1

\usepackage{graphicx}
\usepackage{microtype}
\usepackage[normalem]{ulem}
\usepackage[table,dvipsnames]{xcolor}
\usepackage[colorlinks,urlcolor=NavyBlue,citecolor=NavyBlue,
  linkcolor=NavyBlue,pdfusetitle]{hyperref}
\usepackage[all]{hypcap}

\usepackage{array}
\usepackage{soul}

\renewcommand{\Re}{\text{Re}}
\renewcommand{\Im}{\text{Im}}

\usepackage{orcidlink}

\newcommand{\Cornell}{\affiliation{Cornell Center for
    Astrophysics and Planetary
    Science, Cornell University, Ithaca, New York 14853, USA}}
\newcommand{\CornellPhysics}{\affiliation{Department of Physics, Cornell University, Ithaca, NY, 14853, USA}}
\newcommand{\AEI}{\affiliation{Max Planck Institute for Gravitational Physics (Albert Einstein Institute), Am M\"uhlenberg 1, D-14476 Potsdam, Germany}}
\newcommand{\Caltech}{\affiliation{Theoretical
    Astrophysics 350-17, California
    Institute of Technology, Pasadena, CA 91125, USA}}

\newcommand{\Yukawa}{\affiliation{Center for Gravitational
    Physics and Quantum Information (CGPQI), Yukawa
    Institute for Theoretical Physics
    (YITP), Kyoto University, 606-8502, Kyoto, Japan}}
\newcommand{\Hakubi}{\affiliation{The Hakubi Center
    for Advanced Research, Kyoto University, Yoshida
    Ushinomiyacho, Sakyo-ku, Kyoto 606-8501, Japan}}
\newcommand{\Riken}{\affiliation{RIKEN iTHEMS, Wako,
    Saitama, 351-0198, Japan}}

\newcommand{\Perimeter}{\affiliation{Perimeter Institute for
    Theoretical Physics, Waterloo, ON N2L2Y5, Canada}}

\begin{document}

\preprint{YITP-24-154, RIKEN-iTHEMS-Report-24}
	
\title{Overtones and Nonlinearities in Binary Black Hole Ringdowns}
	
\author{Matthew Giesler
  \orcidlink{0000-0003-2300-893X}}
\email{mgiesler@tapir.caltech.edu}
\Cornell
\author{Sizheng Ma
  \orcidlink{0000-0002-4645-453X}}
\Perimeter
\author{Keefe Mitman
  \orcidlink{0000-0003-0276-3856}}
\Caltech
\author{Naritaka Oshita
  \orcidlink{0000-0002-8799-1382}}
\Yukawa
\Hakubi
\Riken
\author{Saul A. Teukolsky
  \orcidlink{0000-0001-9765-4526}}
\Cornell
\Caltech
\author{\\Michael Boyle \orcidlink{0000-0002-5075-5116}} \Cornell
\author{Nils Deppe \orcidlink{0000-0003-4557-4115}} \Cornell\CornellPhysics
\author{Lawrence E.~Kidder \orcidlink{0000-0001-5392-7342}} \Cornell
\author{Jordan Moxon \orcidlink{0000-0001-9891-8677}} \Caltech
\author{Kyle C.~Nelli \orcidlink{0000-0003-2426-8768}} \Caltech
\author{Harald P. Pfeiffer \orcidlink{0000-0001-9288-519X}} \AEI
\author{Mark A. Scheel \orcidlink{0000-0001-6656-9134}} \Caltech
\author{William Throwe \orcidlink{0000-0001-5059-4378}} \Cornell
\author{Nils L. Vu \orcidlink{0000-0002-5767-3949}} \Caltech

\hypersetup{pdfauthor={Giesler et al.}}
	
\date{\today}

\begin{abstract}
Using high-accuracy numerical relativity waveforms, we confirm the
presence of numerous overtones of the $\ell=2$, $m=2$ quasinormal mode
early in the ringdown of binary black hole mergers. We do this by
demonstrating the stability of the mode amplitudes at different fit
times, ruling out the possibility that a linear superposition of modes
unphysically fits a highly nonlinear part of the waveform. We also
find a number of previously unidentified subdominant second-order
quasinormal modes in the $(2,2)$ mode. Even though these modes are
mathematically nonlinear, they nevertheless confirm the validity of
perturbation theory as a good approximation for describing much of the
ringdown.
\end{abstract}
\maketitle

\section{Introduction}\label{sec:intro}

When two black holes orbit each other, they radiate energy via
gravitational waves and eventually merge to form a single
black hole. The gravitational waveform
rises to a maximum amplitude and then settles
down in a ringdown phase as the remnant black hole sheds its
distortions and evolves toward a final state of equilibrium. Close
enough to equilibrium, the final black hole can be approximated by
linear perturbation theory of the Kerr metric~\cite{Teukolsky:1973ha}.
In this approximation, the radiation is dominated by a
set of quasinormal modes (QNMs).

The QNMs are damped oscillations of the remnant black hole, with
characteristic frequencies determined by the mass and the spin of the
remnant.  Each QNM frequency, $\omega_{\ell m n}$, is complex valued,
encoding the oscillation frequency and damping time. Here
$\ell$ and $m$ label the angular harmonic of a particular
mode. The final index $n$ of $\omega_{\ell m n}$ is the overtone
number. It labels the eigenvalues of the radial equation for the
perturbation.
While the $n=0$ mode, or fundamental mode, was typically of primary
interest in ringdown studies, the overtones, $n>0$, were often
neglected and considered subdominant.  Recent
work~\cite{Giesler:2019uxc} challenged this notion, demonstrating that
overtones are important for modeling binary black hole ringdowns and
that the inclusion of overtones seems to allow for accurate modeling of the
$(2,2)$ component of the waveform as early as its peak.  This result
turned out to be controversial and was met with claims of overfitting
and some doubts about whether the overtones are in fact
physical~\cite{Mourier:2020mwa,Jaramillo:2020tuu,Forteza:2021wfq,Baibhav:2023clw,Cheung:2023vki,Zhu:2023mzv,Nee:2023osy}.
A more in-depth discussion of the connection between this previous
work~\cite{Giesler:2019uxc} and the current work can be found in
Sec.~\ref{sec:prev_work}.

An important issue is how to distinguish a linear superposition
of modes that is physical from a superposition that merely provides
a good fit, that is, overfitting. A good criterion is that if the
modes are really there, their amplitudes should remain constant as the start
time of the fit is varied, apart from the exponential decay of each
mode with its expected decay constant.
Part of the reason for skepticism
in the papers cited above is that they could not demonstrate
that the amplitudes were constant, especially at early times and
for higher overtones. We do so here.

In this paper, we investigate the presence of linear QNMs,
including overtones, and second-order
QNMs~\cite{London:2014cma,Mitman:2022qdl,Cheung:2022rbm,Ma:2022wpv} in the
$(2,2)$ component of the ringdowns of high-accuracy
numerical relativity (NR) simulations.
Our primary focus is a theoretical
understanding of the ringdown, with a goal of
determining the contributions of overtones
and second-order QNMs to the ringdowns of binary
black holes. We aim to determine the physical
picture of the ringdown and how much of it
is described by perturbation theory, either first or second order.
Whether the modes
we identify in this work can be detected in practice with
a certain detector or sensitivity is a separate
question that should be addressed in future work.

Throughout this work, we focus on two
particular NR simulations from the
Simulating eXtreme Spacetimes (SXS) collaboration.
Both simulations are equal-mass aligned-spin binaries
and both are extremely accurate.
The high accuracy of these waveforms is
due to improved methods in extracting the
waveforms from the numerical
evolutions.
In particular, we use
waveforms that have been extracted using a Cauchy evolution via
SpECTRE's Cauchy-characteristic evolution
(CCE) module~\cite{Moxon:2020gha,Moxon:2021gbv,spectre}.
This enables us to not only
obtain much higher-accuracy waveforms, but
because CCE also computes the Weyl scalars, we can map the NR system
to the superrest frame of the remnant black
hole~\cite{MaganaZertuche:2021syq,Mitman:2024uss} using
Bondi-van der Burg-Metzner-Sachs
(BMS) frame fixing~\cite{Mitman:2021xkq,Mitman:2022kwt,Mitman:2024uss}.
This last step is important for removing
mode-mixed QNMs that arise from being in the wrong BMS frame.
If these spurious modes are
not removed but left
unaddressed, they effectively
increase the noise floor and make fitting highly challenging. 
Additional details regarding the NR waveforms and
pre-processing can be found in Sec.~\ref{sec:waveformdata}.

Beyond this, an additional improvement over previous fitting
techniques is the use of
a much more robust nonlinear least
squares algorithm called variable projection. We make use of this algorithm
to perform agnostic searches for QNMs within
the ringdown, identifying the best-fitting
complex frequencies of unknown QNMs. For a detailed
discussion of variable projection,
see Sec.~\ref{sec:dampedsinusoids}.

Identifying and extracting individual QNMs from
the ringdowns of NR waveforms is a challenging
task.
To highlight some of the subtleties
associated with fitting many modes in the
presence of noise, we first treat a simplified
analytic case and discuss the complications
that arise. In particular, we show how sensitive
the highly damped modes are to small amounts
of simulated numerical noise.
In Sec.~\ref{sec:fitting} we introduce
a technique for handling the most sensitive
modes and show how this allows for
better resolution of mode amplitudes
as a function of time.
We use this same technique in Sec.~\ref{sec:results}
to uncover
the many QNMs in the ringdowns of our
two highly accurate NR waveforms.
These analyses confirm that the overtones
are physical and that they can be
stably extracted from binary black
hole ringdowns. Moreover, we also
show a number of previously
unidentified second-order QNMs present
in the dominant $(\ell=2,m=2)$ component of the
ringdown. While the second-order modes
are generally subdominant, they are
in principle detectable, and could
serve as unique probes of the nonlinear
aspects of binary black hole mergers.
Lastly, we explore the accuracy of
using QNMs to predict remnant parameters
and discuss the connection between unmodeled
nonlinearities and biased estimation.
A note on our conventions: we take $c=G=1$ and use the Moreschi-Boyle
conventions of
Refs.~\cite{Boyle2016,Mitman:2021xkq,Mitman:2022kwt,Iozzo:2021vnq,Moreschi:1988pc,Moreschi:1998mw,Dain:2000lij}.

\section{Previous work}\label{sec:prev_work}

In previous work~\cite{Giesler:2019uxc}, we explored the contribution
of overtones to binary black hole ringdowns.
There we showed that the inclusion of overtones
provides an accurate estimate of the underlying remnant mass and spin
as early as the peak\footnote{In this paper, the peak always
refers to the peak amplitude of the (2,2) mode of the strain. The time
of this peak can differ significantly from that of other mode amplitudes,
or of energy fluxes, or of quantities like $\Psi_4$.}
of the strain.  Ref.~\cite{Giesler:2019uxc} also
showed that a model with overtones provides an excellent description
for the post-peak signal, with residuals comparable to the numerical
error in the waveforms used in that study.
Showing that
amplitudes are constant up to the expected decay rates,
a much stricter requirement for
proving the presence of overtones, proved to be more difficult and
Ref.~\cite{Giesler:2019uxc} could only reliably extract up to $n=2$
(with rough evidence for $n=3$ and $n=4$).

In this work, we find that with improved high-accuracy waveforms we
can reliably extract the numerical decay of overtones well beyond
$n=2$.  These improved waveforms, along with improved fitting
techniques, yield residuals at the $10^{-8}$ level, whereas the prior
work had numerical noise at a significantly
higher level of $\sim 10^{-5}.$ This lower noise floor allows us to
identify previously unidentifiable modes and improves our ability to
numerically extract individual modes over a much wider range of
amplitudes and time.

However, under the stricter test of constant mode amplitudes, while
the model still provides accurate predictions for the remnant
parameters and excellent residuals at the peak, some of the modes cannot
be consistently extracted at times very close to the peak.  An
unmodeled, unidentifiable contribution to the waveform in the
immediate post-peak region seems to spoil the stability of the modes
when attempting to extract them numerically.  The time at which we can
show constant mode amplitudes appears to depend on the remnant spin,
as the higher spin system requires a longer wait time.  As can be seen
in Sec.~\ref{sec:results}, stable amplitudes are not achievable until
${}\sim4M$ after the peak in a moderately spinning case and ${}\sim8M$
post-peak in a high-spin case.\footnote{We note that these times
correspond roughly to the times of the peak luminosity, over all
$(\ell,m)$, for each system.}

There are multiple factors that may contribute to our inability to
show constant mode amplitudes at early times in the ringdown.  These
include the prompt response~\cite{Leaver:1986gd,Nollert:1999ji},
evolving QNM amplitudes~\cite{Andersson:1996cm,Chavda:2024awq}, and
effects not describable as perturbations of a single remnant black
hole.  In addition, recent works~\cite{Zhu:2024dyl,May:2024rrg} showed
that changes in the mass and spin of a perturbed black hole generate a
nonlinear response in the QNMs early in the ringdown.  Interestingly,
Refs.~\cite{Zhu:2024dyl,May:2024rrg} showed that despite this
nonlinear response, the resulting waveform is still well-modeled by a
linear superposition of QNMs, which settle into constant amplitudes
very early on.  While Refs.~\cite{Zhu:2024dyl,May:2024rrg} considered
a perturbed single black hole, it is plausible that this effect also
exists in binary mergers and that it might contribute to some of the
unmodeled content in the early ringdown.  Nevertheless, one of the
main conclusions of this paper is that empirically all of these
possible effects make small contributions to the strain observed at
infinity.

It is worth emphasizing that the times referenced above refer
specifically to the time at which we can show all mode amplitudes to
be relatively stable.  While the stability of many modes is influenced
by the unmodeled content at earlier times, the $n\le2$ overtone
amplitudes measured at the peak agree with the amplitudes recovered at
the abovementioned times for each system.  This suggests that these
modes may still be identifiable in an analysis carried out at the
peak.  In Ref.~\cite{Giesler:2019uxc}, a Bayesian analysis exploring
overtones in a LIGO-like detector showed that for realistic
signal-to-noise ratios, a model with just the first few overtones at
the peak is sufficient to accurately recover the remnant mass and
spin.  This is because the first few overtones can be resolved at
higher levels of noise than the higher overtones.  Determining the
optimal time and the set of modes necessary for properly analyzing
ringdown data for different systems and noise levels is left to future
work.  We discuss the issue of start time and the effect of early
unmodeled contributions on the accuracy of our QNM model more in
Sec.~\ref{sec:eps}.

\section{Quasi-Normal Modes}
\label{sec:QNMs}

Linear perturbations of the Kerr metric describing a rotating black
hole can be readily determined by solving a wave-type equation for a
complex scalar $\Psi$ by separation of
variables~\cite{Teukolsky:1972my,Teukolsky:1973ha}. The solution from
doing so can be decomposed into multipoles $\Psi_{\ell m}$ by using
spin-weighted spheroidal
harmonics~\cite{Teukolsky:1972my,Teukolsky:1973ha,Press:1973zz}:
\begin{equation}
\Psi_{\ell m}\sim\int d\omega\,e^{-i\omega t}\sum_{\ell m}
R_{\ell m}(r,\omega).
\label{eq:perts}
\end{equation}
Here the modes $R_{\ell m}$ are a solution of a radial equation
analogous to a quantum mechanical radial equation with some effective
potential.  At large $r$, $\Psi$ encodes the two independent
polarizations of the gravitational wave $h$.  The ringdown solution we
seek for $R_{\ell m}$ describes radiation leaving the domain both at
infinity and at the surface of the black hole---an eigenvalue problem
for frequency $\omega$.  Since energy is being dissipated, the
eigenfrequencies $\omega_{\ell mn}$ are complex. The index $n$ labels
the radial eigenvalues, with $n=0$ corresponding to the least-damped
mode, $n=1$ the next least damped, and so on. Modes for $n>0$ are
called overtones, although they have very different behavior than
overtones in other physical systems. At late times, the
$1/r$ piece of Eq.~\eqref{eq:perts} at large $r$ becomes an outgoing
wave
\begin{equation}
\Psi_{\ell m}=\sum_{\ell mn} C_{\ell mn}e^{-i\omega_{\ell mn}(t-r_*)},
\end{equation}
where $r_*$ is a radial ``tortoise'' coordinate.

The gravitational wave strain $h$ measured by a detector
consists of two polarization states. These are usually
combined into a single complex strain that is
decomposed into modes using spin-weighted spherical harmonics. We will
initially ignore the distinction between spherical and
spheroidal harmonics, so the strain modes at a fixed large $r_*$ are
\begin{equation}
\label{eq:model}
h_{\ell m}=\sum_{n=0}^{\infty} C_{\ell mn}e^{-i\omega_{\ell mn}t}.
\end{equation}
In Eq.~\eqref{eq:model} we have renormalized the amplitudes $C_{\ell
  mn}$ in going from $\Psi$ to $h$.  Each term in this sum represents
a damped sinusoid and gives the mathematical description of a single
QNM with indices $(\ell,m,n)$.
Note that $C_{\ell m n}$ refers specifically to the complex-valued
amplitudes associated with each mode and
that we use $A_{\ell m n} \equiv |C_{\ell m n}|$ to
denote the absolute value of the mode amplitudes.

\subsection{QNM Notation}
The Schwarzschild metric is symmetric under the transformation
$t\to-t$. Accordingly, the QNMs for a given $(\ell,m)$
come in pairs with positive and negative
oscillation frequencies:  $\omega_{\ell m n} = -\omega_{\ell m n}^*$.
The negative-frequency mode is called a mirror mode, since it is
a reflection across the imaginary frequency axis of the positive-frequency
mode. For the Kerr metric, the symmetry transformation is
$t\to -t, \phi \to -\phi$. The QNMs still come in pairs, but the
symmetry relation is now $\omega_{\ell m n} = -\omega_{\ell -m n}^*$.
Thus it is the positive-frequency mode for $-m$ that generates
the negative-frequency partner for the original mode.
Note that because of these relations, tables of QNM frequencies typically
only give the positive-frequency values.

QNMs can also be described as prograde or retrograde. A prograde mode
has $m$ and the real part of $\omega$ with the same sign, so with
the dependence $\exp(-i\omega t + im\phi)$ the pattern speed is in the
direction of increasing $\phi$.

The amplitudes of the QNMs depend on how the modes are excited.
For typical binary black hole mergers, such as those considered
in this paper, the angular momentum in the system ensures that
the prograde modes are excited more strongly than the corresponding
retrograde modes.

In this paper, we will use the notation $(l,m,n)$ to denote a QNM,
adhering to the convention that its real part is always positive. For
example, $(2,2,0)$ corresponds to the prograde fundamental mode
$e^{-i\omega_{220}t}$, while $-(2,-2,0)^*$ represents a retrograde
mode $e^{i\omega^*_{2,-2,0}t}$. The values of $\omega_{220}$ and
$\omega_{2,-2,0}$ are taken from the typical tables of QNM
frequencies. Appendix \ref{appendix:QNM_Notations} gives additional
details about QNM notation.

\subsection{Fitting Damped Sinusoids}
\label{sec:dampedsinusoids}

A key feature of the work reported here is that we do \emph{not} use
standard nonlinear least-squares algorithms to fit QNMs to our
numerical waveforms.
Fitting a sum of damped sinusoids to
data is a tricky problem.  Note that a single damped exponential can
be written in several equivalent forms:
\begin{align}
\label{eq:amp_phase}
Ae^{-i(\omega t +\phi)}&=Ae^{-i\phi}e^{-i\omega t}=Ze^{-i\omega t},\\
Z&= Ae^{-i\phi}\equiv x+iy.
\end{align}
Here $A$ and $\phi$ are real, while $\omega$ can be complex. If
$\omega$ is known, then using the first expression in
Eq.~\eqref{eq:amp_phase} requires nonlinear least squares to fit for
the phase $\phi$, whereas the last expression in that equation makes
clear that the problem actually requires only linear least squares
fitting. However, fitting for the frequencies, whether real or
complex, is inherently a nonlinear least squares problem and generally
much more difficult.

The next important point is that fitting a sum of exponentials with
unknown frequencies or decay constants is numerically ill-conditioned.
This fact has been known for a long time (see, e.g., the 1956 book by
Lanczos~\cite{lanczos1956}), and is continually rediscovered by
physicists. In practice, this means that fitting with the wrong number
of exponentials or with the correct number but noisy data, can lead to
results with large errors.  This sensitivity implies that one needs to
use as robust a method as possible to do the fitting.

The ``best'' algorithm known has also been known for a long time,
since 1973~\cite{golub1973}. It takes advantage of the fact that some
of the parameters in the fit enter the model linearly while others are
nonlinear. Such problems are called separable least squares, and the
algorithm is called Variable Projection, implemented originally in a
Fortran code called \texttt{VARPRO}~\cite{golub1972}.  The idea is to
start with initial guesses only for the nonlinear parameters. Then
standard linear least squares solves for the linear parameters by the usual
analytic process. Next, an iterative nonlinear fitting routine updates
the nonlinear parameters with the linear parameters held fixed. The
whole procedure is then iterated until a suitable tolerance is achieved. The
clever part of the algorithm deals with the Jacobian of the cost function
with respect to the nonlinear parameters that is used in the nonlinear
fitting. This Jacobian has a dependence on the linear parameters because the
nonlinear parameters depend implicitly on the linear
ones. Ref.~\cite{golub1973} worked out this contribution to the
Jacobian---it can be computed explicitly from the analytic solution of the
linear least-squares
problem using linear algebra techniques. In general, this
algorithm is never worse than brute-force nonlinear least squares
fitting, and often succeeds when brute force fails.

Although most of the main results we present in Sec.~\ref{sec:results}
are obtained from simple linear least-squares fits, \texttt{VARPRO} is
crucial in arriving at these results.  The ringdowns we analyze are
composed of an initially unknown number of overlapping individual
quasinormal modes with unknown frequencies and damping times.  While
\texttt{VARPRO} is the best algorithm we know of to tackle this
problem, it is still challenging to simultaneously identify many modes
in the presence of numerical noise when the number of modes and their
frequencies are unknown.
To simplify the problem, we rely on the fact
that many modes decay at different rates and that the majority of the
modes will have decayed below the noise floor at late times.  Given
this, we initially employ \texttt{VARPRO} at very late times where
only one or two long-lived modes are still present and allow
\texttt{VARPRO} to identify modes agnostically, determining the
complex frequencies and complex amplitudes of the slowest decaying
modes.\footnote{Note that we use varying time intervals to identify
stable new modes that show up as we go back in time. These intervals
are longer at later times and shorter at earlier times, when multiple
new modes with rapid decay times appear.}
Once we are confident in the presence and stability of these
modes, we move to a slightly earlier time and look for the next
slowest decaying modes.  To further simplify the fitting, as we step
back to an earlier time and agnostically search for additional modes,
we no longer allow the frequencies of previously found modes to vary.
Instead, we keep them fixed at the analytic values corresponding to
the remnant mass and spin.  This reduces the complexity of the fit to
one or two nonlinear searches at a time, while previously found modes
enter into the fit linearly.  We repeat this process of adding found
modes and nonlinearly searching for new modes until \texttt{VARPRO}
can no longer find any stable modes.

Finally, in regards to initial guesses, 
when the number of modes is reasonable 
and \texttt{VARPRO} is well-behaved, it is insensitive to 
the initial guess when searching for new modes.
Given this, we set the initial guess for the real 
part randomly from the interval $[0,1)$ and 
the imaginary part from an interval of $(-1,0]$,
which are practical values for the ringdown 
frequencies of the waveforms considered in this work.
While \texttt{VARPRO} is capable of imposing bounds, 
we set the bounds for the real and imaginary 
components to be $(-\infty,\infty)$, effectively 
allowing the search to be unbounded.

In this work, we have relied on a modern implementation of
\texttt{VARPRO}~\cite{oleary2013} in Matlab. We have translated this
code into Python so that it can use the nonlinear solvers available in
Scipy. This Python version is publicly available at \cite{varpro}.

\section{Methods}
\label{sec:methods}

\subsection{Waveform Data}
\label{sec:waveformdata}

For our comprehensive ringdown analyses, we use two binary black hole
simulations~\cite{giesler_2025_15086488}, whose inspiral and remnant
parameters can be found in
Table~\ref{table:sims}. The Cauchy evolution of these simulations is
performed using the SpEC code~\cite{SpECCode}, while the waveform and
the Weyl scalars are computed at future null infinity by running the
SpECTRE code's Cauchy-characteristic evolution (CCE)
code~\cite{Moxon:2020gha,Moxon:2021gbv,spectre}.  We then map the
waveform data at future null infinity to the superrest frame of the
remnant black hole $250M$ past the luminosity peak using the package
\texttt{scri}~\cite{scri,Boyle:2013nka,Boyle:2014ioa,Mitman:2021xkq,MaganaZertuche:2021syq,Mitman:2022kwt}. We
stress that this component of the post-processing that fixes the BMS
frame is absolutely crucial for performing robust and accurate
ringdown analyses. By mapping the asymptotic data to the superrest
frame of the remnant black hole, one ensures that the remnant is at
the origin, at rest, has its spin aligned with the $\hat{z}$-axis, and
is not supertranslated relative to the usual Kerr metric. Thus, there
can be no unexpected mode mixing between the spherical harmonic modes,
which would complicate the QNM fits, as was found in, e.g.,
Refs.~\cite{Ma:2022wpv,Cheung:2023vki}.

\setlength{\tabcolsep}{6pt}
\begin{table}[]
	\centering
	\begin{tabular}{c|c|c|c|c|c}
		SXS ID & $q$ & $\chi_{1z}$ & $\chi_{2z}$ & $M_{f}$ & $\chi_{f}$\\
		\hline \hline \vspace{-8pt}\\
		BBH:2420 & 1.0 & 0.2 & 0.2 & 0.95 & 0.75 \\
		BBH:2423 & 1.0 & 0.85 & 0.85 & 0.91 & 0.92 \\
	\end{tabular}
	\caption{\label{table:sims} Inspiral and remnant parameters for
          the two binary black hole simulations used in our
          ringdown analyses.
          Here $q$ is the mass ratio and
          $\chi_z$ is the
          $z$-component of the BH spin.
          In both simulations, the
          binaries are of equal mass and the BH spins
          are aligned with the
          orbital angular momentum. The remnant
          parameters are the final mass $M_f$
          and the final BH spin $\chi_f$.
          }
\end{table}

Once the strain and the Weyl scalars are obtained at future null
infinity in the superrest frame of the remnant black hole,
we utilize \texttt{scri} to perform
one more post-processing step that simplifies the ringdown analyses.
As outlined in Eq. (12) of Ref.~\cite{Mitman:2020bjf} as well as
elsewhere in
Refs.~\cite{Geroch:1977big,Ashtekar:1981bq,Ashtekar:2019viz}, because
of the supertranslation symmetry at future null infinity, the strain
obeys the supermomentum balance law:
\begin{align}
\label{eq:supermomentumbalancelaw}
h=\frac{1}{2}\bar{\eth}^{2}\mathfrak{D}^{-1}\left[-\left(\Psi_{2}+\frac{1}{4}\dot{h}\bar{h}\right)+\frac{1}{4}\int_{-\infty}^{u}|\dot{h}|^{2}du\right].
\end{align}
The notation here is that of Ref.~\cite{Mitman:2020bjf} and is not
important for this paper, other than to note that $h$ is the strain.
The first term in Eq.~\eqref{eq:supermomentumbalancelaw} is related to
the Bondi mass aspect and is effectively a combination of the mass and
the current multipole moments. The second term in
Eq.~\eqref{eq:supermomentumbalancelaw} corresponds to an energy flux
and is often interpreted as the source of the null
memory~\cite{Zeldovich:1974gvh,Braginsky:1985vlg,Braginsky:1987kwo,Payne:1983rrr,Christodoulou:1991cr,Blanchet:1992br,PhysRevD.45.520,Flanagan:2015pxa,Mitman:2020pbt,Mitman:2020bjf}.
It is a purely nonlinear contribution in the sense that it depends on
the strain quadratically. Consequently, when fitting the strain with
linear QNMs, this term raises the effective noise floor, as it can
only be modeled with second-order QNMs.\footnote{In particular, this
term measures the null memory sourced by QNMs and should be dominated
by the second-order QNM $(2,2,0)-(2,2,0)^{*}$ for typical binary black
hole mergers.} Because of this, rather than fitting the full strain,
which includes this nonlinear contribution, we instead only fit the
contribution to the strain coming from the first term in
Eq.~\eqref{eq:supermomentumbalancelaw}.
Although this term will still
have nonlinear QNM contributions that we will have to fit, by
neglecting the contribution from the null memory the fits should be
more straightforward to carry out.
Finally, because of residual
supertranslations that may exist from the BMS frame fixing being
imperfect, instead of fitting the strain (which can be shifted by a
constant), we fit the news, i.e., the first time derivative of the
strain.

\begin{figure}[t]
  \includegraphics[width=1.0\columnwidth]{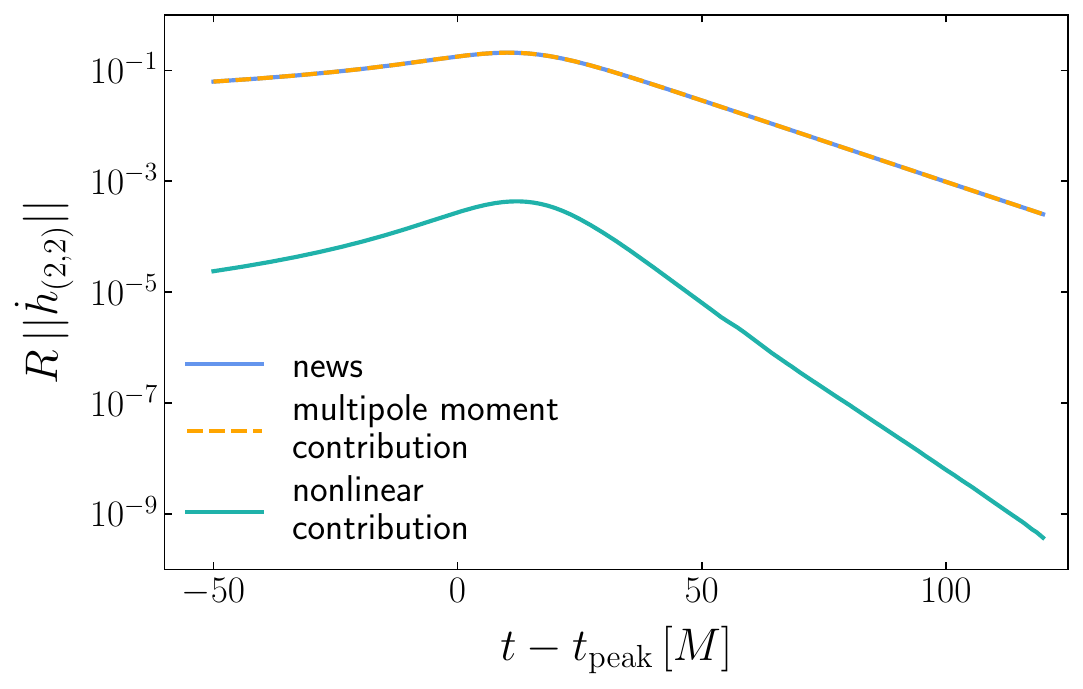}
  \caption{The $(2,2)$ component of the news and
    the two terms of Eq.~\eqref{eq:supermomentumbalancelaw}
    that contribute to the news for SXS:BBH:2423.
    The news and the contribution from the
    multipole moment, the first term
    in Eq.~\eqref{eq:supermomentumbalancelaw},
    overlap and are effectively indistinguishable.
    The nonlinear contribution, which is a few orders
    of magnitude smaller than the news, 
    is also shown. This nonlinear contribution
    refers to the energy flux described in
    Sec.~\ref{sec:waveformdata} and corresponds
    to the second term in Eq.~\eqref{eq:supermomentumbalancelaw}.}
  \label{fig:news_vs_mass_aspect}
\end{figure}

To help clarify this point regarding the nonlinear term, in
Fig.~\ref{fig:news_vs_mass_aspect} we show the $(2,2)$ mode of the
news as well as the contributions from the multipole moments and the
nonlinear term in Eq.~\eqref{eq:supermomentumbalancelaw}. As is
clearly illustrated, the nonlinear term in
Eq.~\eqref{eq:supermomentumbalancelaw} is three orders of
magnitude smaller than the full news.  Consequently, since this term
would simply detract from our ability to resolve overtones by
including certain extra, nonlinear contributions, we do not include it
in the waveforms that we fit in this work.

Finally, the last step in preparing the waveforms to be fit
with QNMs is to adjust the scale of the time coordinate (and hence
the derived frequencies)
to match the convention in tabulated values.
This step is unfortunately often ignored in the literature, but
is key for performing robust and accurate black hole spectroscopy.
In particular, SpEC uses the total Christodoulou mass of
the binary as the mass unit,
whereas tables set the mass unit to unity.
Thus the time coordinate of SpEC waveforms must be multiplied by
the total Christodoulou mass.
Even though this mass is typically within $10^{-4}$ of unity,
failing to make this correction can strongly bias QNM fits.

\subsection{Fitting}\label{sec:fitting}

\begin{figure*}[t]
  \includegraphics[width=1.9\columnwidth]{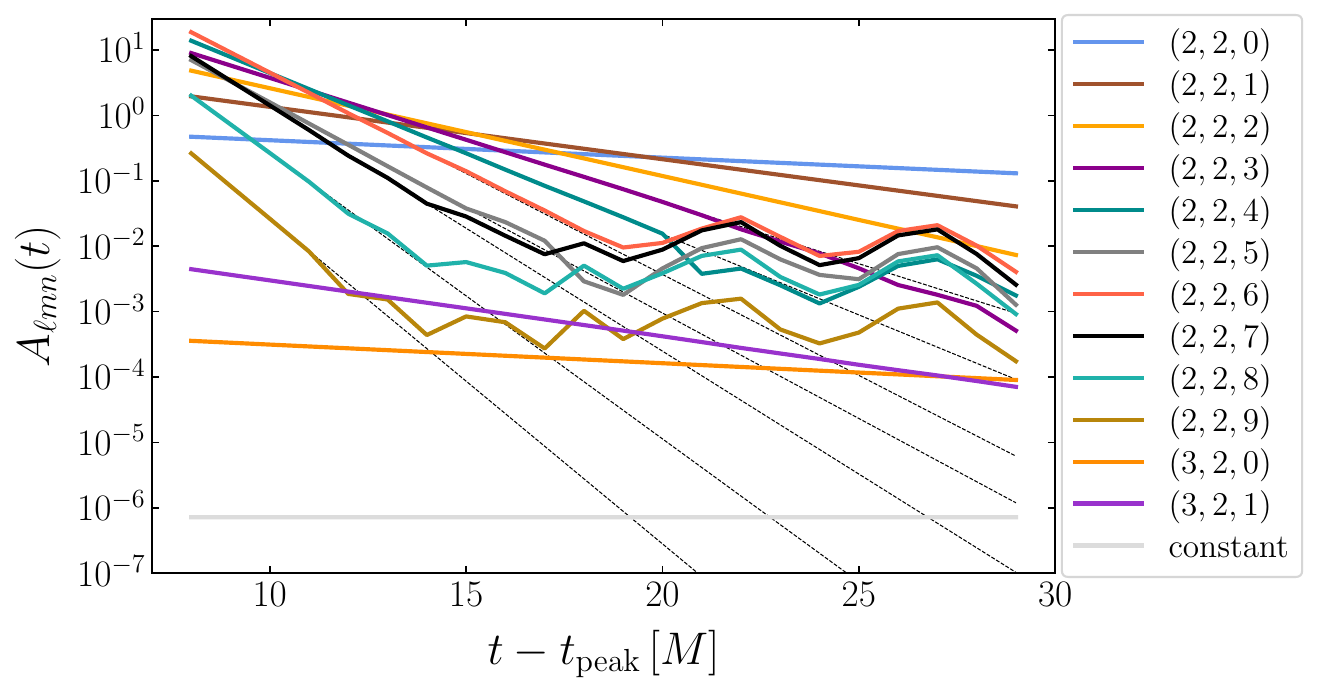}
  \caption{Mode amplitudes as a function of time, recovered
    from a linear least-squares fit to
    an analytic waveform in the presence of
    Gaussian noise at a level of $\sim 10^{-8}$.
    The modes included are various $(2,2,n)$ and $(3,2,n)$ modes
    with amplitudes and phases modeled from SXS:BBH:2423, as
    well as a constant term with an amplitude
    of $\sim 5\times10^{-7}$. The recovered amplitudes
    are the solid colored lines, while the black dotted lines
    represent the analytic expectations for each mode.
    Notice that once the $(2,2,9)$ mode begins to deviate
    from its analytic value because of the small amount of
    added noise,
    other highly damped modes eventually follow suit.
    This extreme sensitivity to low-level noise is one of the
    primary challenges
    to stably resolving highly damped QNMs.
  }
  \label{fig:2423_analytic}
\end{figure*}

Identifying the modes present in the ringdown of an arbitrary
numerical waveform comes with a number of challenges.
To illustrate these challenges, we
first consider an idealized analytic ringdown model.
We construct an analytic waveform modeled
after the ringdown of SXS:BBH:2423 and its remnant
parameters. The analytic model is a superposition of
modes, as in Eq.~\eqref{eq:model}, including the
$(2,2,n)$ modes with $n=\{0...9\}$, the $(3,2,0)$ mode and
its first overtone, the $(3,2,1)$ mode. In addition
to these damped sinusoids, we also add a constant
term
to help stabilize our fits, since such a term seems to be present
in our numerical relativity waveforms for reasons that are not clear.
The constant we fit tends to be three orders of magnitude smaller
than the smallest damped sinusoid, which emphasizes the ill-conditioned
nature of the fitting.
The frequencies, $\omega_{\ell m n}$,
are determined by the remnant of SXS:BBH:2423 and
the complex amplitudes, $C_{\ell mn}$, are consistent
with those we find below in SXS:BBH:2423.

In such a purely analytic model, we can use linear
least squares with varying start times to fit for the
modes we have included. We recover the expected
amplitudes as they decay over time until they
eventually reach roundoff. However,
in realistic numerical waveforms, numerical noise
interferes with the ability to resolve the
individual modes below some level.
In order to better understand the role of numerical
noise, we add a small amount of Gaussian noise to
the analytic model.
In this particular example, we add Gaussian noise
with an average amplitude of $\sim 2\times 10^{-8}$.
For context, the numerical noise often present in
SXS extrapolated waveforms is closer to $10^{-4}$ -- $10^{-5}$,
so the amount of noise added in this example is orders
of magnitude smaller than that present in publicly
available waveforms.

The results of this procedure are shown in Fig.~\ref{fig:2423_analytic}.
Notice that while the least damped modes are well resolved
across a wide range of times, the highest damped modes
are most sensitive to the tiny amount of noise
we have added.
In Fig.~\ref{fig:2423_analytic}, the $n=9$ and $n=8$
modes are the first to be affected by the added noise,
deviating from their expected decays at around $12$ -- $13M.$
Although the noise is only around $10^{-8}$, these
modes become unresolvable at much larger amplitudes,
$\sim 10^{-3}$ for $n=9$ and $\sim 10^{-2}$ for $n=8.$
Once the higher damped modes begin to deviate,
the
next highest damped modes are eventually affected
and the deviations cascade through the lower $n$,
yielding inaccurate amplitudes for many of the
modes.
This high sensitivity to low noise levels even
in analytic models means
that highly damped modes are likely to be very difficult
to resolve in even the best numerical
waveforms publicly available,
\emph{and even when they are really present.}

The early deviation from the expected decay
that occurs for some of the mid-range $n$, say $n=5$,
is in part due to the higher modes, $n > 5$,
becoming unresolvable yet remaining in the set
of modes we use to fit to the data.
For example, consider Fig.~\ref{fig:2423_analytic},
where we are fitting up to $N=9$, where $N$ denotes the
maximum overtone number $n$ used in a fit.
We see how the $n=5$
mode begins to deviate from its expected decay
at around $16M$, when its amplitude is still $10^{-2}$.
This is well above the amplitude for which we
are able to resolve the highest damped mode, $n=9$.
One remedy for this issue is to discard
the highest modes once they are affected by
the noise, which in turns extends our
ability to resolve some of the lower modes.
In practice, this means that once we reach
a time where the $n=9$ mode can no longer
be resolved for a given noise level, the
set of $(2,2)$ modes that we include in the
fit is limited to $N=8$ rather than $N=9$.
Similarly, once we reach a time where the
$n=8$ mode has reached its limit, we drop
that mode and limit the set to $N=7$.

\begin{figure}[ht]
  \includegraphics[width=1.0\columnwidth]{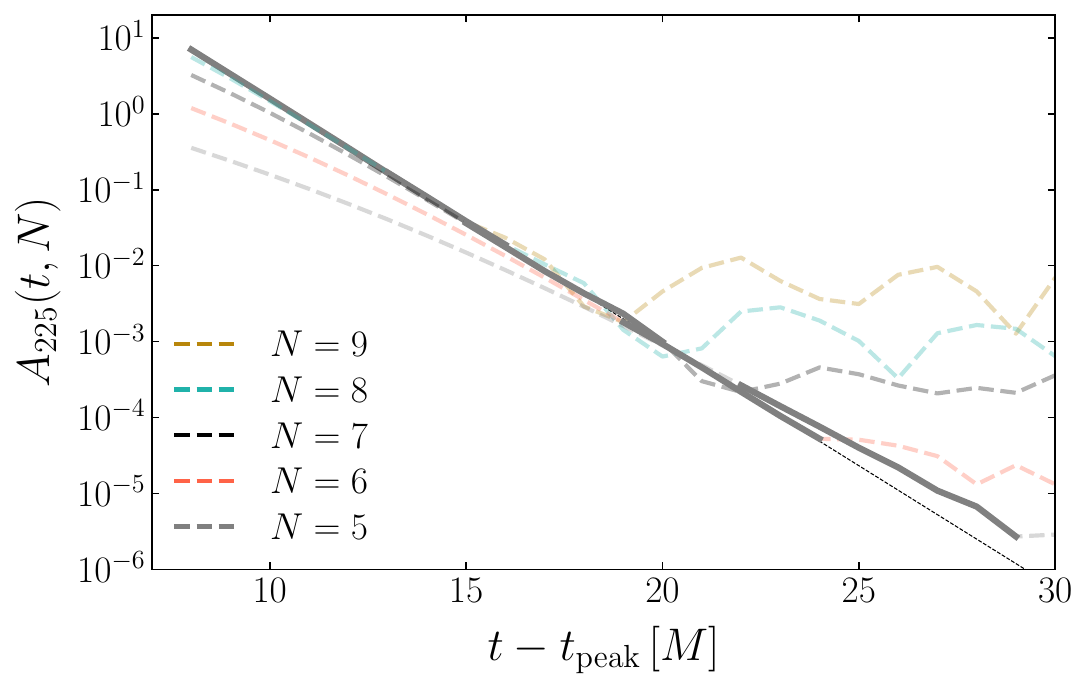}
  \caption{
    The $(2,2,5)$ mode amplitude recovered from a linear
    least-squares fit to the same analytic waveform shown
    in Fig.~\ref{fig:2423_analytic}. The solid gray line
    is a composite amplitude of the $n=5$ mode obtained
    through the process of dropping higher modes once
    they become unstable due to noise.
    Each dashed curve shows the $n=5$ amplitude recovered
    from a fit including up to $N$ overtones of the
    $(2,2,n)$ QNMs. Notice that the curve associated
    with $N=9$ is the first to deviate from the expected
    decay and that models with a lower $N$ at later
    times are in better agreement with the analytic expectation.
    This dropping of modes results in improved
    mode amplitude extraction, extending the
    amplitude down a few orders of magnitude
    and out an additional $\sim 10M$ compared
    to just fitting with $N=9$.
   }
  \label{fig:2423_n_decay_analytic}
\end{figure}

In Fig.~\ref{fig:2423_n_decay_analytic},
we show how this process of dropping higher modes
leads to better resolvability of lower modes,
and in this particular case, the $n=5$ mode.
There are five different curves
in Fig.~\ref{fig:2423_n_decay_analytic},
each corresponding to the amplitude of the $n=5$ mode when
the fit is limited to just $N$ overtones
of the $(2,2)$ mode. The $N=9$ curve is
identical to the $n=5$ curve of
Fig.~\ref{fig:2423_analytic}, as that fit
includes up to $N=9$ overtones.
Notice in Fig.~\ref{fig:2423_n_decay_analytic}
that each time $N$ is reduced,
the $n=5$ mode can be resolved at a
lower amplitude than the previous $N$.
To distinguish the regions where
certain $N$ are valid, each dashed
curve is overdrawn with a solid gray
curve when it is the preferred set.
Once the fitting process with mode dropping
is complete, we are left with a composite
$n=5$ amplitude that now extends down
to roughly $10^{-5}$ and out to about
$29M$. So, through the process of dropping
modes, we extend our ability to resolve
this particular mode by three orders of
magnitude in the amplitude and increase
the time over which it is resolved by
$\sim10M$. A direct comparison can
be made by comparing the gray composite
amplitude with the dashed $N=9$ curve,
which corresponds to the best resolvability of
$n=5$ in the absence of mode dropping.

Although we focused here on the $n=5$ mode
of the analytic example, the process of dropping
modes leads to better resolvability of all modes.
This is because including modes in the fit that
are no longer present in the data spoils the
resolvability of those modes still present.
We find through this analytic example
that the process of dropping modes
is critical to properly resolving modes.
This example also emphasizes the ill-conditioned
nature of fitting damped sinusoids to data.
In the next section, we use this same
technique to properly resolve the modes of
numerical relativity waveforms.

\section{Results}\label{sec:results}

Having described our methods for identifying modes
with \texttt{VARPRO} in Sec.~\ref{sec:dampedsinusoids}
and our mode-dropping technique in the previous section,
we now turn to identifying and fitting the QNMs of
two high-accuracy numerical relativity waveforms.
In particular, we study one highly spinning remnant,
SXS:BBH:2423, and one more moderately spinning remnant,
SXS:BBH:2420. The binary and remnant parameters for
each simulation are listed in Table~\ref{table:sims}.
For each waveform, we use \texttt{VARPRO}
to identify the QNMs present in the ringdowns.
We do this by starting at late times and working
our way backward, letting \texttt{VARPRO} identify
the best-fitting frequency and damping times at
each point in time. Each time we find a stable
frequency and damping time that corresponds to
a QNM associated with the asymptotic remnant
parameters, that QNM is permanently added to
the fitting set. Previously found modes then
enter into the fit linearly alongside a
nonlinear search for additional modes
through \texttt{VARPRO}. We continue this
process until no additional stable modes
are found.
With this procedure, we are able to
identify a large set of stable QNMs in
both simulations. These modes include a large
number of
overtones, multiple modes from spherical-spheroidal
mixing, and a number of second-order QNMs.

Note that there are two distinct fitting procedures used to obtain the
following results.  The first procedure involves nonlinear fitting
using \texttt{VARPRO} to identify the QNMs in a particular waveform,
as described above.  The second fitting procedure is a linear least
squares method coupled with a mode-dropping technique to show the
stability of the QNMs identified through \texttt{VARPRO} in the first
step.  This second procedure is critical to show the stability of QNM
amplitudes in the presence of numerical noise as highlighted in the
simplified analytic case within Sec.~\ref{sec:fitting}.

\begin{figure*}[t]
  \includegraphics[width=2.0\columnwidth]{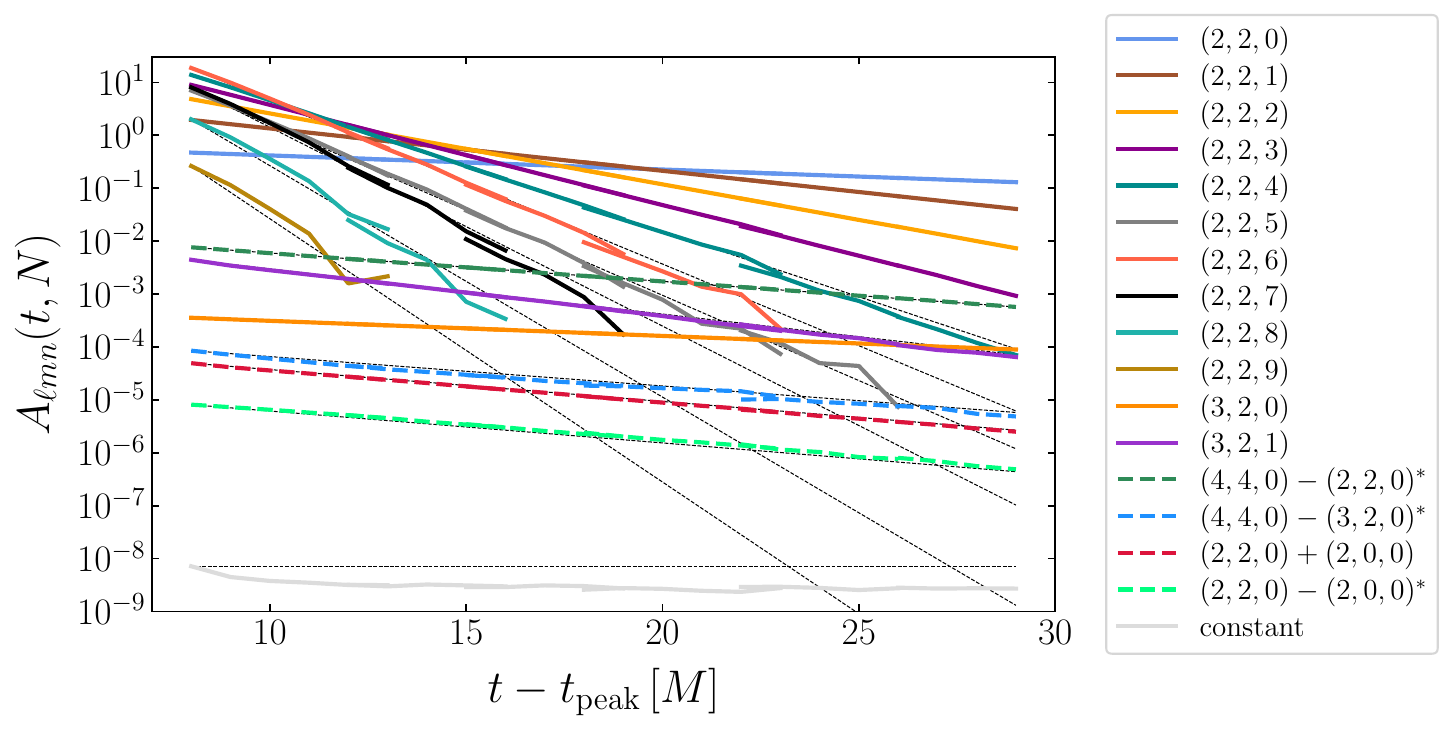}
  \caption{
    QNM amplitudes as a function of time
    from linear least-squares fits of the $(2,2)$ component of SXS:BBH:2423.
    The linear QNMs are denoted by solid lines while the second-order
    QNMs are indicated by dashed lines. Here again, the dotted black lines
    represent the expected analytic decay of each mode.
    The discontinuities in the
    amplitudes are due to the mode-dropping technique that was shown
    to improve mode extraction in the analytic model considered
    in Sec.~\ref{sec:fitting}.
    At each point where a QNM's amplitude
    ends discontinuously, this means that mode has been dropped
    from the set of modes in the fit. This is the case for
    the highest-damped modes, which are more easily
    affected by the numerical noise present in the NR waveforms.
    Aside from these modes, which eventually get dropped,
    the remaining QNMs are well resolved over a significant
    window of time. The majority of the
    second-order QNMs are subdominant to all linear contributions,
    except for the $\omega_{440} - \omega_{220}^{*}$ QNM.
    This particular QNM, with its larger amplitude
    and relatively slow decay, is a good candidate
    for probing nonlinear contributions to
    ringdowns in future gravitational wave observations.
    }
  \label{fig:2423_amplitudes}
\end{figure*}

\subsection{High-spin NR Ringdown}\label{sec:HSring}

The first case we consider is SXS:BBH:2423,
in which the QNMs decay more slowly because of
its more rapidly spinning remnant.
This property of high-spin remnants means
that the more rapidly decaying overtones
are longer lived and therefore more easily
resolvable.
We show
a numerical decomposition of the ringdown
of SXS:BBH:2423 into its individual modes
in Fig.~\ref{fig:2423_amplitudes}.
We see that the ringdown
contains a multitude of QNMs,
including the fundamental $(2,2,0)$
mode and nine of its overtones, the
$(3,2,0)$ mode and one of its overtones,
and four second-order QNMs.
For each mode, we include a dotted line
indicating the expected analytic decay,
which is the path a stable well-resolved
mode should follow.
The discontinuities in the amplitudes of the
individual modes are a consequence of the
mode dropping technique we employ,
which is necessary because of the numerical
noise present and was discussed in
more detail in Sec.~\ref{sec:fitting}.

A surprising aspect of this result
is the presence of four previously
unidentified second-order QNMs, which
are all exceptionally stable.
A second-order QNM, or quadratic QNM,
is generated from the product
of two linear QNMs.
So its frequency is a sum or difference:
\begin{equation*}
  \omega = \omega_{\ell_{1} m_{1} n_{1}} + \omega_{\ell_{2} m_{2} n_{2}} \quad \text{or} \;
  \quad \omega = \omega_{\ell_{1} m_{1} n_{1}} - \omega_{\ell_{2} m_{2} n_{2}}^{*} \, .
  \label{eq:so}
\end{equation*}
These would lead to modes with
$m = m_{1} + m_{2}$ and $m = m_{1} - m_{2}$.
The modes we find that satisfy $m=2$ in this ringdown are
$\omega_{440} - \omega_{220}^{*}$,
$\omega_{440} - \omega_{320}^{*}$,
$\omega_{220} + \omega_{200}$,
and $\omega_{220} -\omega_{200}^{*}$.
All of these modes are denoted by dashed lines in
Fig.~\ref{fig:2423_amplitudes}.
While the real part of the second-order QNM frequency is the
sum or difference of the real parts of the linear QNMs,
in both of the above cases, the imaginary part of the
second-order QNM is always the sum of the imaginary
components of the linear QNMs. Thus, second-order
QNMs decay more quickly than the linear QNMs they are
generated from. Consequently, as can be seen
in Fig.~\ref{fig:2423_amplitudes},
the second-order QNMs decay more quickly than the
fundamental mode, but more slowly than the overtones.
This property of quadratic QNMs means that they
occupy regions of the complex plane that make
them reasonably easy to identify when relying on
\texttt{VARPRO} to find modes.
Moreover, while the second-order QNMs are subdominant
to the overtones at early times, some of these
modes remain measurable in the ringdown after
most of the overtones have decayed away.
This means that even at fairly late times in
the ringdown, for a system like SXS:BBH:2423,
the fundamental remains accompanied by
the $(2,2,1)$, the $(3,2,0)$, and the
$\omega_{440} - \omega_{220}^{*}$ second-order QNM.
This means that, even at late times, a
multimodal analyses of the ringdown
may well be required with future high-accuracy detectors.

It is important to note that the data
in Fig.~\ref{fig:2423_amplitudes} begins
at $t = t_{\mathrm{peak}} + 8M$.
This is because, before this time, \texttt{VARPRO}
can no longer reliably identify any additional
stable complex frequencies. There appears
to be some additional content in this region
of the ringdown, but it is apparently not
well-modeled by a damped sinusoid with a
fixed frequency. Consequently, performing
a decomposition before this time results
in unstable mode amplitudes, especially
for the highest-damped modes, which are
extremely sensitive to small amounts of
noise or unmodeled data.
One possible issue related to stably
decomposing modes earlier than $8M$
is that some of the $(\ell,m)$'s
contributing to second-order QNMs do
not peak until after the peak of the
$(2,2)$. For instance, the
$\omega_{440} - \omega_{220}^{*}$ QNM
is sourced by the $(4,4,0)$ mode,
yet the $(4,4)$ contribution to the
strain does not peak until $\sim 7.5M$
after $t_{\mathrm{peak}}$.
In fact, the average peak flux
over the two-sphere occurs $\sim 10.5M$
after the peak of the $(2,2)$, which
indicates some second-order contributions
are likely still being sourced and
are not well modeled by a constant
exponential decay.

\begin{figure}[h]
  \includegraphics[width=1.0\columnwidth]{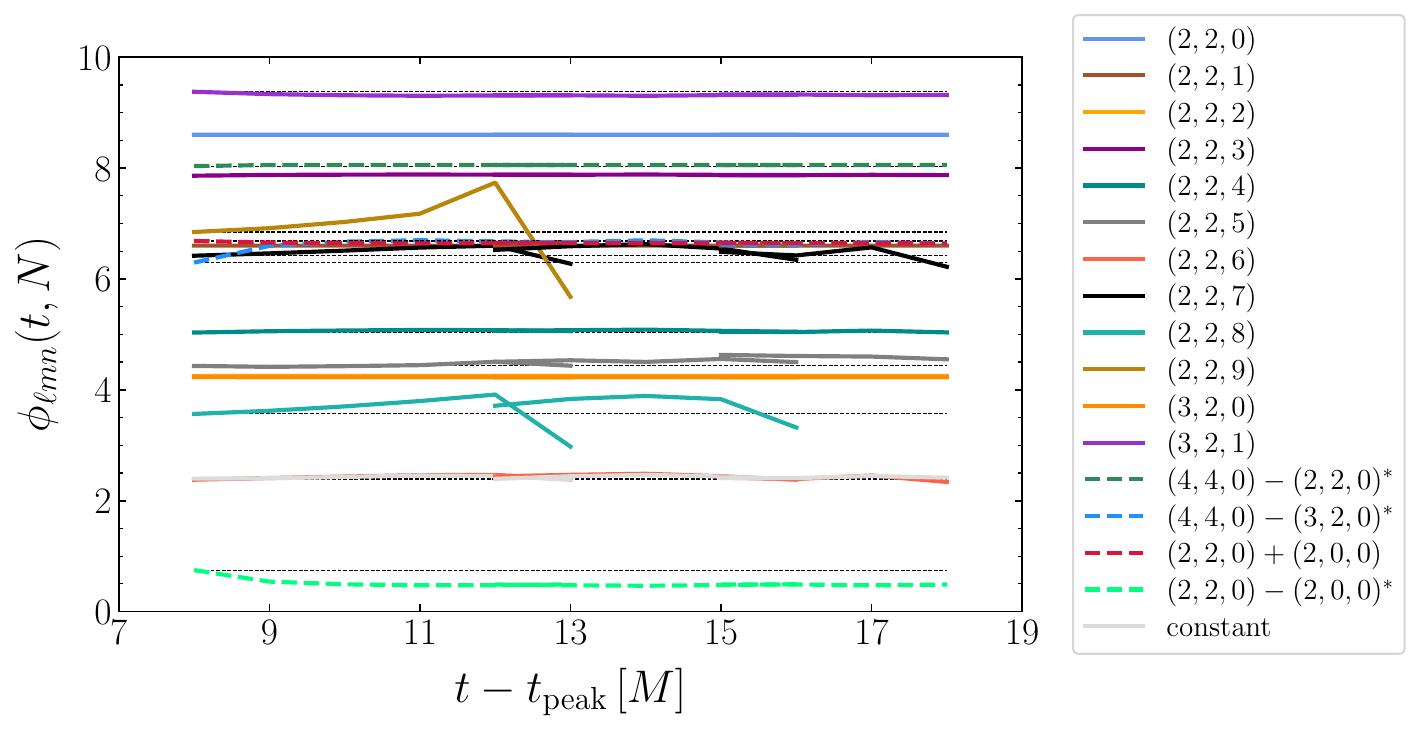}
  \caption{
    QNM phases as a function of time
    from linear least-squares fits of the $(2,2)$
    component of SXS:BBH:2423.
    The phases are obtained from the fits shown
    in Fig.~\ref{fig:2423_amplitudes} and the
    line styles here are consistent with
    that caption. The discontinuities in the phases
    of the higher damped QNMs
    are caused by the mode dropping described
    in the caption of Fig.~\ref{fig:2423_amplitudes}.
    The phases are
    generally constant, with the higher overtones
    deviating first because of their sensitivity
    to numerical noise. We have not mapped the phases
    back to a conventional
    $2\pi$ range
    to improve distinguishability.
  }
  \label{fig:2423_phases}
\end{figure}

Finally, one additional check on the stability of
modes is to check the phases of each QNM,
as these should remain constant
over time. In Fig.~\ref{fig:2423_phases},
we show the phase of each mode over a
window of $10M$, where it is apparent
that the most stable modes have constant phases
throughout. As was the case with the amplitudes,
the most quickly decaying overtones are
the first to become unstable, which
can be seen in the deviation of the
phases from their initial values.

\subsection{High-spin n=5 QNM}
\label{sec:n5}

\begin{figure}[]
  \includegraphics[width=1.0\columnwidth]{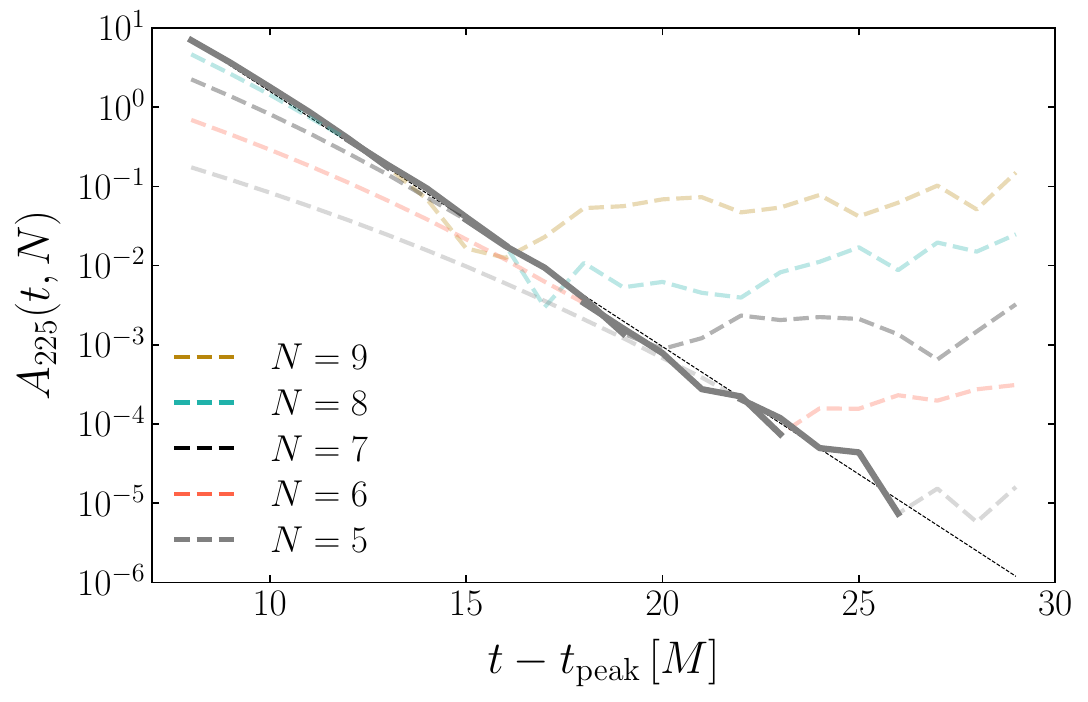}
  \caption{The $(2,2,5)$ mode amplitude recovered from linear
    least-squares fits to the $(2,2)$ mode of SXS:BBH:2423.  The solid
    gray line is a composite amplitude of the $n=5$ mode obtained
    through the mode dropping process.  Each dashed curve shows the
    $n=5$ amplitude recovered from a fit including up to $N$ overtones
    of the $(2,2,n)$ QNMs.  This result is consistent with the example
    of extracting higher $n$ from an analytic ringdown shown in
    Fig.~\ref{fig:2423_n_decay_analytic} and described in
    Sec.~\ref{sec:fitting}.  The recovered amplitude is in good
    agreement with the expected analytic decay (black dotted line),
    spanning roughly five orders of magnitude in the
    amplitude over a range of $15M$.}
  \label{fig:2423_n5_decay}
\end{figure}

In Fig.~\ref{fig:2423_n5_decay} we show
the stability of the $n=5$ mode in the ringdown of SXS:BBH:2423,
a particularly challenging mode,
which we can resolve over nearly $20M$ and across more
than five orders of magnitude in the amplitude,
showing consistency with the expected decay for a remnant
with $\chi=0.92$. We highlight this mode because of its
high overtone number and its anomalous behavior in the
set of $(2,2)$ QNMs. This particular mode exhibits peculiar
behavior with respect to the other $n$ in its frequency
and excitation factors for very high spins.
In Appendix~\ref{appendix:exc_facs}, we
discuss this mode further and explore the
relationship between mode amplitudes and
excitation factors.

\begin{figure*}[t]
  \includegraphics[width=2.0\columnwidth]{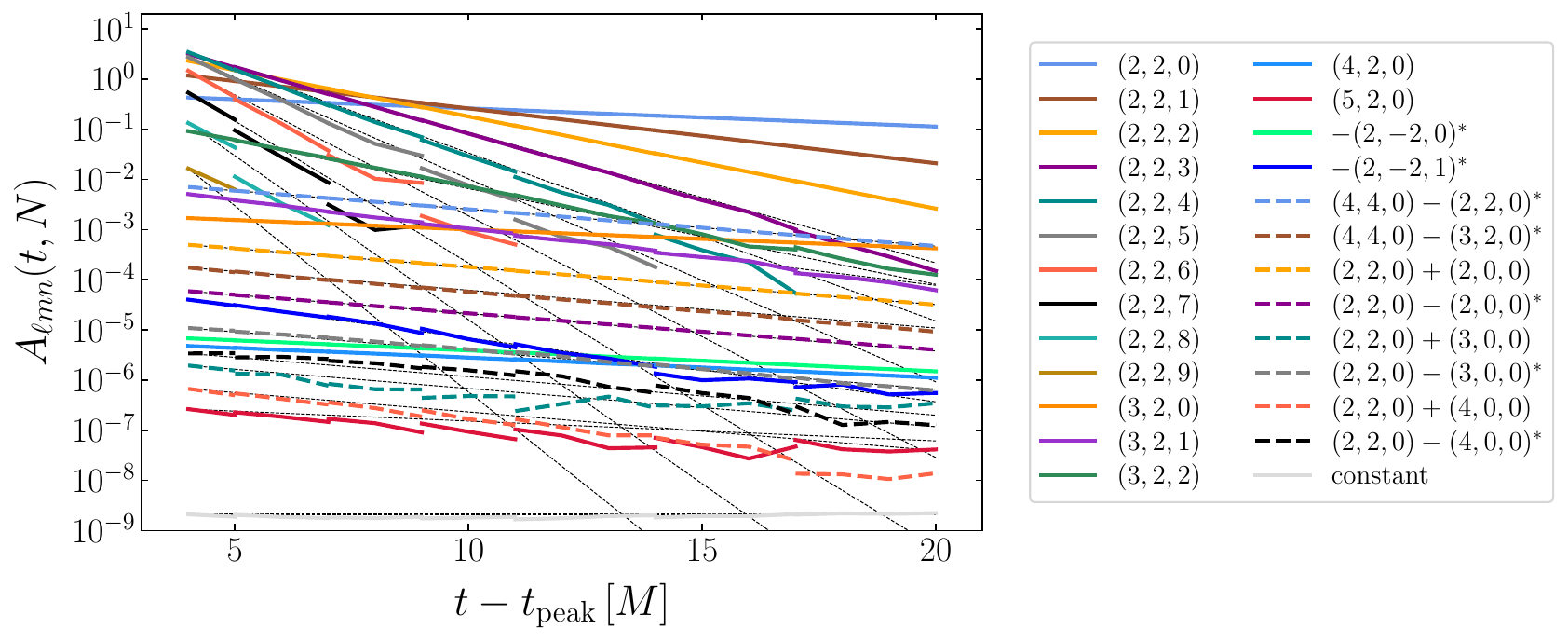}
  \caption{
    QNM amplitudes as a function of time
    from linear least-squares fits of the $(2,2)$ component of SXS:BBH:2420.
    The linear QNMs are denoted by solid lines while the second-order
    QNMs are indicated by dashed lines. The dotted black lines
    again represent the expected analytic decay of each mode.
    As detailed in the caption of Fig.~\ref{fig:2423_amplitudes},
    the high-spin case,
    the discontinuities in the
    amplitudes are caused by dropping certain fast-decaying QNMs.
    The amplitudes of the highest overtones
    are more difficult to resolve in this lower-spin
    case because of the increased number of modes and
    the increased damping that comes with lower spins.
    Additionally, the lowest-amplitude modes are
    also more challenging to resolve, especially
    as they decay further into the numerical noise.
    Aside from this, the majority of the first- and
    second-order QNMs are well resolved.
    Note that there are a number of additional
    modes present in this simulation as compared
    to the higher-spin case, including some
    retrograde modes, higher overtones of the $(3,2)$,
    and four additional second-order QNMs.
    Here again, the $\omega_{440} - \omega_{220}^{*}$ QNM
    is the dominant second-order mode, indicating that this
    mode in particular may be an ideal candidate for
    future gravitational wave observations.
  }
  \label{fig:2420_amplitudes}
\end{figure*}

\subsection{Moderate-spin NR Ringdown}\label{sec:MSring}
Next, we consider SXS:BBH:2420, a system with a more
moderately spinning remnant, specifically $\chi=0.75$.
This remnant spin is more representative of typical binary mergers
and is closer to the remnant spin of a GW150914-like
merger.
Since the spin is lower, the QNMs decay more quickly than those
of the system we considered in Sec.~\ref{sec:HSring}.

Employing the same methods used in Sec.~\ref{sec:HSring},
including the mode dropping method of Sec.~\ref{sec:fitting},
we present a numerical decomposition of the
QNMs in the ringdown of SXS:BBH:2420 in
Fig.~\ref{fig:2420_amplitudes}.
What immediately stands out here in comparison
to Fig.~\ref{fig:2423_amplitudes} is a significant
increase in the number of modes present.
As these simulations have different binary
configurations, the initial conditions at
the onset of the ringdown result in a
different excitation of modes.

As Fig.~\ref{fig:2420_amplitudes} shows,
we find many of the same modes present in
SXS:BBH:2423, namely the $(2,2,0)$ mode
and many overtones, the $(3,2,0)$ QNM and
its first overtone, and the same second-order
QNMs. However, we also find
additional modes that do not appear in the
ringdown of SXS:BBH:2423. These include
a second overtone associated with the $(3,2,0)$
mode, the $(3,2,2)$, and some additional
QNMs from spherical-spheroidal mixing,
the $(4,2,0)$ and $(5,2,0)$.
Interestingly, we also find four additional
second-order QNMs, in particular:
$\omega_{220} + \omega_{300}$,
$\omega_{220} - \omega_{300}^{*}$,
$\omega_{220} + \omega_{400}$,
and $\omega_{220} -\omega_{400}^{*}$.
The final two additional modes present
here are the retrograde QNM
$-\omega_{2,-2,0}^{*}$
and its first overtone
$-\omega_{2,-2,1}^{*}$.

As was the case for SXS:BBH:2420,
many of the $(2,2)$ overtones are well
resolved, aside from the quickest decaying
modes such as the $n\ge7$
QNMs, which are highly susceptible to
small amounts of noise.
These QNMs are also more sensitive
to the dropping of the next highest mode,
leading to larger discrepancies in the
amplitudes measured before and
after the drop.
The $(3,2,n)$ modes are all very well
resolved, including the $(3,2,2)$, which
is the first time this mode has been
found in the $(2,2)$ harmonic.
While the $(4,2,0)$ QNM is easily measurable,
the $(5,2,0)$ mode has a very small amplitude,
making it challenging to cleanly extract.
The two retrograde modes, which are
apparently swallowed up in the higher-spin
case, are both relatively stable.
A Fourier transform of the ringdown signal
confirms that these modes reside on
the opposite side of the frequency axis
relative to the $(2,2,n)$ QNMs.

As for the second-order QNMs, the
highest amplitude modes are as stable
as many of the first order QNMs, but
the lower amplitude modes are somewhat
susceptible to noise.
Notice in Fig.~\ref{fig:2420_amplitudes},
that the $\omega_{440} - \omega_{220}^{*}$
QNM is again the highest amplitude
quadratic mode and of sufficient amplitude
that it will survive until late in the
ringdown, subdominant only to the $(2,2,0)$,
$(2,2,1)$, and $(3,2,0)$.
The three other second-order QNMs,
$\omega_{440} - \omega_{320}^{*}$,
$\omega_{220} + \omega_{200}$,
and $\omega_{220} -\omega_{200}^{*}$
that were also present in SXS:BBH:2423,
have moderate amplitudes that remain stable
throughout. Notice though that while
the $\omega_{440} - \omega_{320}^{*}$ QNM
was larger than the $\omega_{220} + \omega_{200}$ QNM
in the higher spin case, the order of these
two modes is flipped in this ringdown.
The remaining four second-order QNMs,
ordered by amplitude are
$\omega_{220} - \omega_{300}^{*}$,
$\omega_{220} -\omega_{400}^{*}$.
$\omega_{220} + \omega_{300}$,
and $\omega_{220} + \omega_{400}$.
While the first two are relatively
stable throughout, the two at
lower amplitude become noisy as
they decay further into the noise.

The data in Fig.~\ref{fig:2420_amplitudes}
does not start at $t_{\mathrm{peak}}$,
but instead at
$t = t_{\mathrm{peak}} + 4M$.
As was the case with SXS:BBH:2423,
\texttt{VARPRO} cannot identify a
stable frequency before this time.
We suspect this is likely for the same
reasons discussed in Sec.~\ref{sec:HSring},
namely unmodeled content and sourcing
of the second-order
modes. Here, the
$(4,4)$ does not peak until $\sim 4.5M$
after $t_{\mathrm{peak}}$ and its average peak
flux over the two-sphere is $\sim 7M$ after the $(2,2)$ peak.
Once again, stable amplitudes are
achievable roughly around the time
the $(4,4)$ peaks.
Although we suspect this contributes
to the instability of amplitudes early
on, additional nonlinearities may also
play a role here. In the two cases,
the time until stable amplitudes are
achieved appears to follow the usual
relationship between remnant spin
and decay rates, meaning that any
nonlinearities should decay away
more slowly in the higher-spin case
and more quickly in the lower-spin
case, as we see here.

\subsection{Remnant Parameters}\label{sec:eps}
\begin{figure}[b]
  \includegraphics[width=1.0\columnwidth]{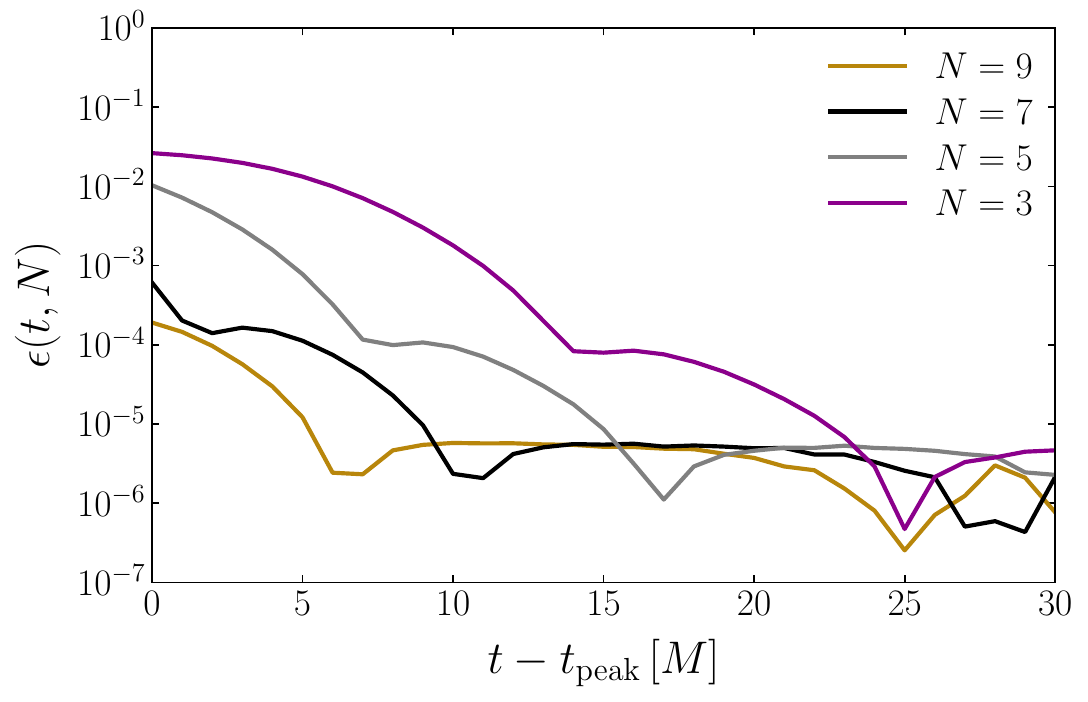}
  \caption{Measure of error $\epsilon$, defined in
    Eq.~\eqref{eq:epsilon}, as a function of time for different $N$.
    This measure of error quantifies the difference between the
    best-fit mass and spin versus the NR asymptotic remnant mass and
    spin for SXS:BBH:2423.  Each fit includes the set of QNMs shown in
    Fig.~\ref{fig:2423_amplitudes}, but with the $(2,2,n)$ modes
    limited to a maximum $N$ overtones.
    There is a floor near $10^{-5}$ that each
    $N$ reaches once that model is preferred.
    The $N=9$ set accurately predicts the remnant quantities
    at a level of $10^{-4}$ at $t_\mathrm{peak}$
    and eventually reaches the $10^{-5}$ floor
    near the
    time at which we can show stable amplitudes in
    Fig.~\ref{fig:2423_amplitudes}. The other $N$ eventually
    reach similar values once the higher overtones excluded
    from those sets have sufficiently decayed away.
  }
  \label{fig:2423_epsilon}
\end{figure}
So far, we have used the NR asymptotic remnant
mass and spin, $M_{f}$ and $\chi_{f}$, to determine
the QNM frequencies, $\omega(M_{f},\chi_{f})$, used
in our fits.
To test whether this is indeed the best-fitting
$M$ and $\chi$, we repeat the fits while allowing
these parameters to vary and inform the QNM
frequencies through $\omega(M,\chi)$.
In the end, the best-fitting $M$ and $\chi$ is the
set that produces the smallest residual---keeping
in mind that the underlying mass and spin as measured
in the NR simulation are unknowns in this process.
To see how the best-fitting $M$ and $\chi$ compare
to the NR quantities, $M_{f}$ and $\chi_{f}$, we
use
\begin{equation}
  \epsilon = \sqrt{(\delta M_f/M)^2 + (\delta \chi_f)^2}
  \label{eq:epsilon}
\end{equation}
as our measure of error,
where $\delta M_f$ and $\delta \chi_f$ are the differences
between the best-fit values and the NR measured quantities.
This allows us to quantify how well the model predicts
the true underlying asymptotic remnant quantities and
to explore the performance of the model at
different times in the ringdown.

\begin{figure}[b]
  \includegraphics[width=1.0\columnwidth]{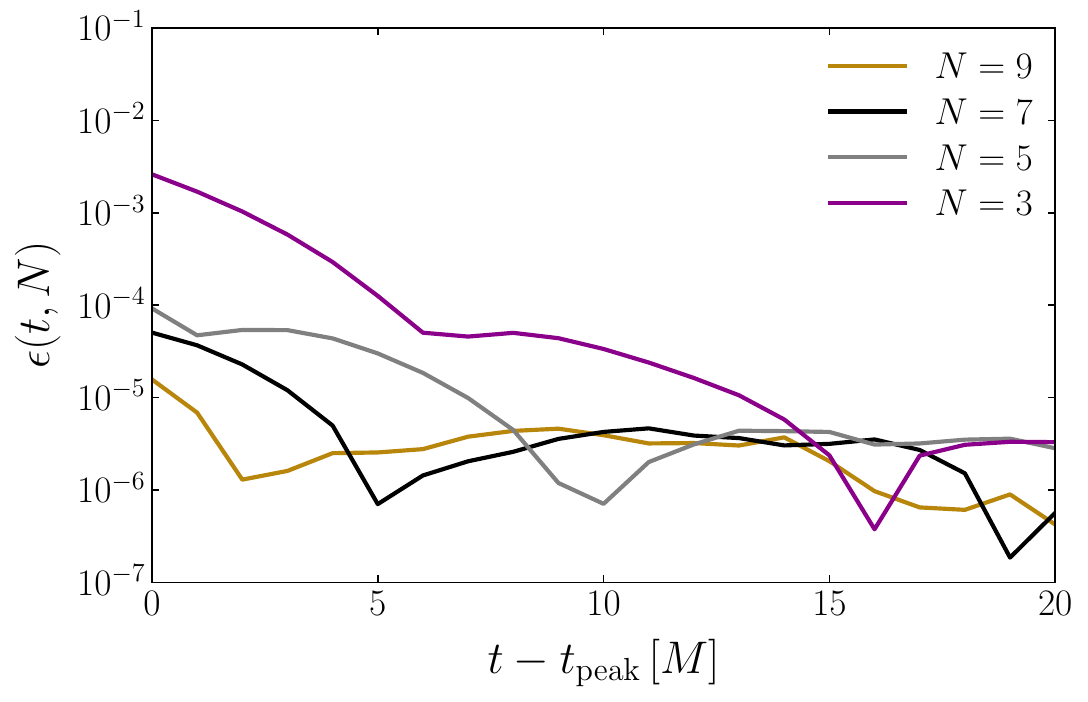}
  \caption{
    Measure of error $\epsilon$, as defined in Eq.~\eqref{eq:epsilon},
    for SXS:BBH:2420.
    As detailed in the caption of Fig.~\ref{fig:2423_epsilon},
    this quantity measures the error in the mass and spin
    preferred by each set of QNMs compared to the
    true NR asymptotic remnant mass and
    spin. Here, as in Fig.~\ref{fig:2423_epsilon}, each
    $N$ reaches a floor below $10^{-5}$ once that model
    is optimal. In this case, each $N$ produce smaller
    values of $\epsilon$ at $t_\mathrm{peak}$ compared
    to those in Fig.~\ref{fig:2423_epsilon},
    which is consistent with finding stable amplitudes
    at an earlier time compared to the higher spin case.
    Similarly, each $N$ reaches the $10^{-5}$ floor
    earlier, since the QNMs for this case are more
    highly damped. Although $N=9$ does not reach
    a minimum until $\sim 4M$, it is worth noting that
    this set of QNMs accurately predicts the
    remnant quantities with an error of $\sim 10^{-5}$
    as early as $t_\mathrm{peak}$.
  }
  \label{fig:2420_epsilon}
\end{figure}

We begin with a focus on the ringdown of the high-spinning
remnant of SXS:BBH:2423.
At each point in time, we fit a set of QNMs to the data and
find the best combination of $M$ and $\chi$ that minimizes the residual.
We then compute the corresponding $\epsilon$ from Eq~\eqref{eq:epsilon}.
In Fig.~\ref{fig:2423_epsilon}, we show $\epsilon$ for
a range of models over a range of
$30M$ starting from $t_\mathrm{peak}$.
All of the modes used in the fits of
Fig.~\ref{fig:2423_amplitudes} are again used here, but we
limit the set of $(2,2,n)$ modes to $n=\{0,...,N\}$,
where $N$ is the maximum $n$ included in the fit.
In Fig.~\ref{fig:2423_epsilon}, we find that when the
appropriate $N$ is used, the model predicts the remnant
quantities remarkably well, with $\epsilon<10^{-5}$.
This small deviation from the NR remnant quantities
indicates how well the QNM model fits the data and
reflects the predictive power of accurate fitting.
Fig.~\ref{fig:2423_epsilon} also allows us to explore
the shortcomings of a given model by observing
when $\epsilon$ deviates from the minimum achievable
values. For instance, considering $N=3$,
we see that $\epsilon$ is as large as $\sim10^{-2}$
at early times and does not achieve $10^{-5}$
until $\sim25M.$ This is because many overtones
present in the data earlier on are not included
in the set used for fitting and it is not until
those modes have decayed away does $N=3$ become
the appropriate model. The results and reasoning
are similar for the other $N$.

While the deviations in $\epsilon$ for $N=\{3,5,7\}$
in this example are understood to be caused by
ignoring overtones present in the data, the
case of $N=9$ is less well understood.
In Fig.~\ref{fig:2423_epsilon}, we see that
for $N=9$, $\epsilon$ stabilizes around $10^{-5}$
about $8$ -- $9M$ after the peak. This is consistent
with our ability to show stable amplitudes
after this time in Fig.~\ref{fig:2423_amplitudes}.
However, before this time, we see $\epsilon$ deviates
from this minimum value and rises to $10^{-4}$
by $t_\mathrm{peak}$.
This again tells us that there is
unmodeled content in this region that we are
not currently including in the fit.
This is consistent with the discussions in
the two previous sections on unstable
amplitudes at early times, where we
speculate this is caused either by extended
sourcing of second-order QNMs
beyond $t_{\mathrm{peak}}$ or by
additional unknown nonlinearities.
Still, it is reassuring
that even without including this unknown
content, the set of modes we are fitting with
remain an excellent model capable of accurately
predicting the remnant quantities at the
level of $10^{-4}$ as early as $t_\mathrm{peak}$.

We repeat this same procedure for the
case of SXS:BBH:2420, with the results
shown in Fig.~\ref{fig:2420_epsilon}.
Here we have included all of the QNMs
shown in Fig.~\ref{fig:2420_amplitudes}
while limiting the set of $(2,2,n)$
modes to $n=\{0,...,N\}$.
The values of $\epsilon$ at $t_{\mathrm{peak}}$
for all $N$ are all smaller than those
of SXS:BBH:2423 in Fig.~\ref{fig:2423_epsilon}.
While the floor for $\epsilon$ is generally
the same for both cases, the $\epsilon$'s are
smaller at $t_{\mathrm{peak}}$ in this case
since our model is stable much closer to
$t_{\mathrm{peak}}$. The $\epsilon$
for $N=9$ is slightly biased at early times,
but reaches its minimum at $\sim 4M$ once
the unmodeled content has decayed away.
This too
is consistent with the time at which we
observe stable amplitudes
in Fig.~\ref{fig:2420_amplitudes}.
So, while we cannot show stable amplitudes
for all QNMs before $t = t_{\mathrm{peak}}+ 4M$,
the model captures the majority of the
ringdown signal, precisely predicting the
remnant parameters down to a few times
$10^{-5}$ as early as $t_{\mathrm{peak}}$.
This is consistent with the result found
previously in Ref.~\cite{Giesler:2019uxc}
for a waveform with similar parameters.

\section{Discussion and conclusions}\label{sec:conclusion}

Understanding the nature of binary black hole ringdowns
is critical to future endeavors probing Einstein's
general theory of relativity. Whether it is 
black hole spectroscopy or searching for black hole
mimickers, it is essential to understand the physical
aspects of ringdown signals.
Using some of the highest-accuracy ringdown waveforms
available from numerical relativity coupled with
a robust nonlinear fitting scheme, we reveal
details of the ringdown that were previously
unachievable.
Through advances in numerical relativity
waveform extraction, namely CCE, BMS
frame fixing, and removing contributions from the memory,
the noise floor in the ringdown
of these waveforms is orders of magnitude
smaller than those previously available. 
This, along with the use of
variable projection (\texttt{VARPRO}),
unveils a number of QNMs in the dominant
quadrupolar $(2,2)$ component of the ringdown 
that were previously undiscovered.
This includes a number of second-order
QNMs and overtones of modes present
through spherical-spheroidal mixing.
In addition to these newly
observed modes, we also improve on our
ability to numerically extract the many overtones
associated with the $(2,2)$ QNM.
Through simplified analytic explorations,
we show the high sensitivity of overtones
to small amounts of noise and provide an
improved method for extending our ability
to show stable amplitudes across time.

In the ringdown of a highly spinning remnant,
SXS:BBH:2423, we find that the overtones
have stable amplitudes over at least
$10M$ up to the $n=7$ overtone, and the
next highest two to a lesser extent.
Beyond this, we also find four second-order
QNMs, the much sought after nonlinearities
that have remained elusive in
binary black hole ringdowns.
These four QNMs ---
$\omega_{440} - \omega_{220}^{*}$,
$\omega_{440} - \omega_{320}^{*}$,
$\omega_{220} + \omega_{200}$,
and $\omega_{220} -\omega_{200}^{*}$
--- turn out to be remarkably stable
alongside the many first-order QNMs. 
While these second-order modes are
generally subdominant in amplitude,
the loudest of these,
$\omega_{440} - \omega_{220}^{*}$,
is likely the best candidate for
future observation and may serve as
an excellent probe of the nonlinear
nature of binary black hole mergers.

In the case of a more typical
moderately spinning remnant, SXS:BBH:2420,
we find an even greater abundance of modes.
Because of the short-lived nature of overtones
at lower spins, the highest overtones
are shown to be stable over a smaller
window of time, while the lower $n$
remain more easily resolvable. 
In addition to finding more modes
from spherical-spheroidal mixing, we
find up to the second overtone of the
$(3,2)$ QNM and four more 
second-order QNMs. These additional
second-order QNMs are subdominant to
the four found in SXS:BBH:2420, coming
from interactions with the $(2,2,0)$
and lower amplitude $(3,0,0)$ and
$(4,0,0)$ QNMs.
The $\omega_{440} - \omega_{220}^{*}$ QNM
is again the largest amplitude quadratic
mode, indicating that this mode likely dominates
the second-order QNMs in aligned-spin binary
black hole mergers like those considered in this work.
Future explorations of more generic
systems are necessary to reveal more
about the spectrum and excitability
of second-order modes in binary ringdowns.

Through this work, we show that with the
proper fitting techniques and high-accuracy waveforms
we can confidently identify and extract the individual
QNMs of the ringdown spectrum.
We also confirm the importance of overtones and their
physical contribution to binary ringdowns.
Further, we show there are indeed nonlinearities
in binary black hole ringdowns. One kind of nonlinearity is
identifiable second-order QNMs. It is worth emphasizing that,
while mathematically these count as a nonlinearity, they
are computable by applying linear perturbation
theory to second order. In addition, there may perhaps be
some still unknown contributions in the early
ringdown, with relatively small amplitude.

The second-order QNMs in binary ringdowns
will undoubtedly be targets of future
detectors probing general relativity.
Further studies are necessary
to determine the detectability of these
modes in future detectors.

\begin{acknowledgments}
This work was supported in part by the Sherman Fairchild Foundation,
by NSF Grants PHY-2207342 and OAC-2209655 at Cornell, and by NSF
Grants PHY-2309211, PHY-2309231, and OAC-2209656 at Caltech.  Research
at Perimeter Institute is supported in part by the Government of
Canada through the Department of Innovation, Science and Economic
Development and by the Province of Ontario through the Ministry of
Colleges and Universities. Support for this work was provided by
NASA through the NASA Hubble Fellowship grant HST-HF2-51562.001-A
awarded by the Space Telescope Science Institute, which is operated
by the Association of Universities for Research in Astronomy,
Incorporated, under NASA contract NAS5-26555.
This work was also supported in part by the Japan Society for the
Promotion of Science (JSPS) KAKENHI Grant No.\ JP23K13111
and the Hakubi project at Kyoto University.

\end{acknowledgments}

\section*{DATA AVAILABILITY}
The data that support the findings of this article are
openly available~\cite{giesler_2025_15086488}.

\appendix

\section{Alternative QNM Notations}
\label{appendix:QNM_Notations}
In this paper, we use the notation $(\ell,m,n)$ to label a QNM, with
the convention that the real part of the frequency is always
positive. This makes it easy to find numerical values for the
QNMs in tables, which typically use this convention.
Alternatively, a QNM can be labeled using four indices
$(\ell,m,n,p)$, where its temporal and azimuthal components are
explicitly given by:
\begin{align}
    e^{-i\omega_{\ell mn}^p t} e^{im\phi}.
\end{align}
The fourth index $p$ can be defined in two different ways:
\subsection{Convention 1}
\label{app:convention_1}
\begin{figure}[h]
  \includegraphics[width=1.0\columnwidth]{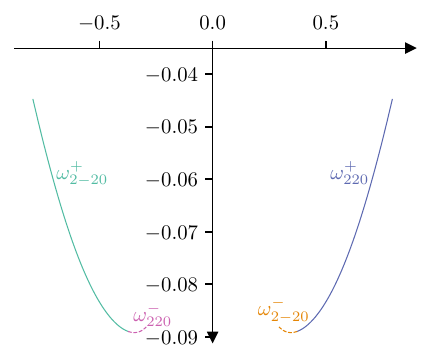}
  \caption{Four branches of QNMs for Kerr BHs under Convention 1,
    where $p =\rm{sgn}(\Re(\omega)) \times \rm{sgn}(m)$.
    Each curve is parameterized by $\chi$ from $0$ to $1$,
    with $\chi=0$ corresponding to the point where the solid
    and dashed lines meet.
    The x-axis corresponds to $\Re(\omega)$ and the y-axis
    to $\Im(\omega)$.
    Note that pairs of mirror modes are symmetric about the imaginary
    axis.
    The prograde modes are denoted by solid lines while the
    retrograde modes are dashed.
  }
  \label{fig:QNM_notation_1}
\end{figure}

We follow \cite{Isi:2021iql} and define
$p =\rm{sgn}(\Re(\omega)) \times \rm{sgn}(m)$.
As an example, Fig.~\ref{fig:QNM_notation_1}
shows the real and imaginary parts of QNMs $(\ell=2,m=\pm 2,n=0,p=\pm)$
for Kerr BHs parameterized by $\chi$.
The four branches
correspond to different signs of $m$ and $p$. The two branches on the left
$(p\times\rm{sgn}(m)<0)$ are mirrors of the branches on the right
$(p\times\rm{sgn}(m)>0)$.
Additionally, a positive $p$ corresponds to a prograde mode
while negative $p$ indicates a retrograde mode.
The Kerr symmetry yields
\begin{align}
  \omega_{\ell mn}^p = -({\omega}_{\ell-mn}^{p})^*.
\end{align}
This notation is related to ours via
\begin{align}
  (\ell,m,n)\to \omega_{\ell mn}^+, \quad -(\ell,m,n)^* \to \omega_{\ell,-m,n}^{{\rm sgn}~m}.
\end{align}

For example, in Fig.~\ref{fig:2420_amplitudes}, the retrograde mode
$-(2,-2,0)^*$ corresponds to $\omega_{2,2,0}^-$. The negative $p$
specifies that it is a retrograde mode, while $p\times m<0$ indicates
that it appears in the negative frequency band. The second order mode
$(4,4,0)-(2,2,0)^*$ represents $\omega_{440}^+\times\omega_{2,-2,0}^+$,
which is generated by two prograde modes. As
shown in Fig.~\ref{fig:QNM_notation_1}, $\omega_{2,-2,0}^+$ and
$\omega_{220}^+$ are symmetric about the imaginary axis, indicating
they are mirrors of each other.

\subsection{Convention 2}
\begin{figure}[h]
  \includegraphics[width=1.0\columnwidth]{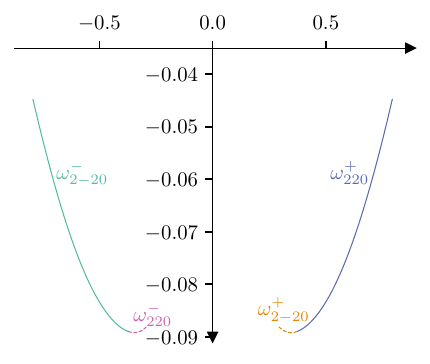}
  \caption{
    The four branches of QNMs for Kerr BHs under Convention 2.
    This is the same as Fig.~\ref{fig:QNM_notation_1},
    but with $p=\rm{sgn}(\Re(\omega))$.}
  \label{fig:QNM_notation_2}
\end{figure}

Convention 1 has a caveat, as it becomes ambiguous when $m=0$. One can
instead define $p =\rm{sgn}(\Re(\omega))$
\cite{MaganaZertuche:2021syq}. In Fig.~\ref{fig:QNM_notation_2}, we
explicitly label the four branches of $(\ell=2,m=\pm 2,n=0,p=\pm)$ for
Kerr BHs, which are again parameterized by $\chi$.
In this case, positive $p$ refers to a standard mode and
negative $p$ refers to a mirror mode.
In this notation, prograde modes satisfy $p\times\rm{sgn}(m)>0$
and retrograde modes have $p\times\rm{sgn}(m)<0$.
The Kerr symmetry now yields
\begin{align}
  \omega_{\ell mn}^p = -({\omega}_{\ell-mn}^{-p})^*.
\end{align}
This second notation is related to ours via
\begin{align}
  (\ell,m,n)\to \omega_{\ell mn}^+, \quad -(\ell,m,n)^* \to \omega_{\ell,-m,n}^{-}.
\end{align}
For example, in Fig.~\ref{fig:2420_amplitudes}, the second order mode
$(2,2,0)-(2,0,0)^*$ refers to $\omega_{2,2,0}^+\times
\omega_{2,0,0}^-$, whereas $(2,2,0)+(3,0,0)$ refers to
$\omega_{2,2,0}^+\times \omega_{3,0,0}^+$.

\section{Excitation Factors}
\label{appendix:exc_facs}
The formalism for extracting the QNM contribution to a waveform was
first studied in detail by Leaver~\cite{Leaver:1986gd}.  Since we are
dealing with a linear wave-type equation, the Fourier transform of the
solution can be written with a Green's function:\footnote{See
Sec.\S7.3 in Ref.~\cite{Morse1953}. A pedagogical example is in
Ref.~\cite{Berti:2006wq}.}
\begin{equation}
\Psi_{\ell m}(\omega,r)\sim\int G(r,r',\omega) T(r',\omega)\,dr'.
\label{eq:green}
\end{equation}
In Eq.~\eqref{eq:green} $T$ represents the source term. This could be,
for example, the stress-energy from a small particle falling into the
black hole.  In our case, it is the initial condition for the problem,
i.e., the gravitational field outside the black hole at the time we
decide to use perturbation theory.  The Green's function $G$ can be
constructed from the two linearly independent solutions to the
homogeneous radial equation.

The solution $\Psi_{\ell m}(t,r)$ in the time domain requires an
inverse Fourier transform of Eq.\ \eqref{eq:green} along a contour in
the complex plane (essentially a Laplace transform). The QNM
frequencies correspond to the poles of $G$ in the complex plane, with
$n$ labeling the poles in order of increasing imaginary part. The
contribution of $G$ to the solution \eqref{eq:green} will include a
sum over the residues $E_{\ell mn}$ at these poles.  So the mode
amplitudes $C_{\ell mn}$ will be a product of two factors,
\begin{equation}
C_{\ell mn}=E_{\ell mn}T_{\ell mn},
\end{equation}
where the $T_{\ell mn}$ come from the radial integral in
Eq.\ \eqref{eq:green} and depend explicitly on the source of the
perturbation.

Now comes a key point: If the source factors $T_{\ell mn}$ are slowly
varying with $n$, then the mode amplitudes $C_{\ell mn}$ will be
nearly proportional to $E_{\ell mn}$.  Since the $E_{\ell mn}$ come
from the homogeneous solutions to the perturbation equation, they are
universal constants that depend only on $M$ and $\chi$---the mass and
dimensionless spin of the remnant black hole---and are called
excitation factors.

There have been several advances in the calculation of
excitation factors since the work of
Leaver~\cite{Leaver:1986gd,Glampedakis:2003dn,Berti:2006wq,Zhang:2013ksa,Oshita:2021iyn}.
Most relevant for us is the work of Oshita~\cite{Oshita:2021iyn},
who showed that
the assumption that the
$T_{\ell  mn}$ do not vary much does appear to be valid for black hole
mergers, at least for the low values of $n$ that we are interested in.

In light of this, along with the difficulties
in individually resolving the
$n=5$ mode in previous work,
one proposed method to determine the presence
of this mode is to fit for the amplitudes of
all $n$ and compare the structure of the measured
amplitudes to the structure of the excitation factors.
In the work of Oshita~\cite{Oshita:2021iyn},
the $n=5$ excitation factor
begins rapidly decreasing with respect to the
other $n$ beyond $\chi\sim0.9$.
This peculiar feature in the excitability of the
modes suggests that the $n=5$ mode should have a reduced
amplitude at higher spins, creating a unique signature
in the structure of the mode amplitudes.
So, if we
look at the amplitudes of the modes as a function of
$n$, a dip should be expected in the amplitude of
the $n=5$ mode once the remnant spin is sufficiently
high.
If the measured $C_{22n}$ agree well with
the known $E_{22n}$, this indicates that the fit
is sensitive to each $n$ within the ringdown.

\begin{figure}[]
  \includegraphics[width=1.0\columnwidth]{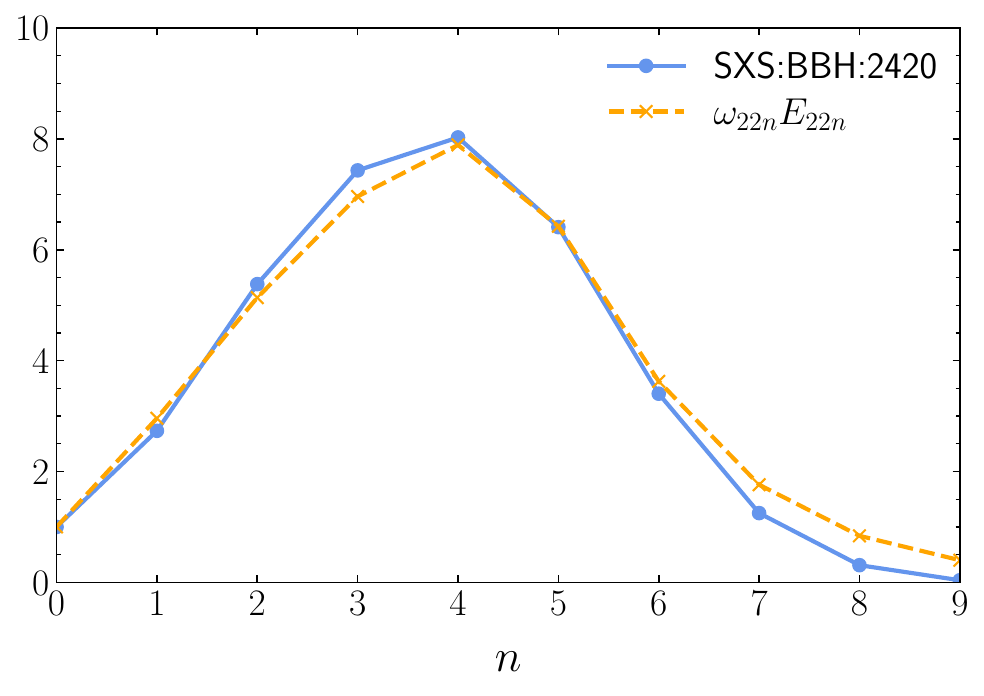}
  \caption{
    The $(2,2,n)$ NR amplitudes as measured in the
    moderate-spin case of SXS:BBH:2420 and the excitation
    factors $\omega_{22n} E_{22n}$ for $\chi=0.75$.
    The excitation factors are computed using $t_0 = 2.5$
    and both curves are normalized such that
    $C_{220} = \omega_{220}E_{220} = 1$.
  }
  \label{fig:2420_exc_facs}
\end{figure}

But, before we consider this unique feature of highly spinning
remnants, it is useful to first consider how well the excitation
factors $E_{22n}$ agree with the measured NR amplitudes $C_{22n}$ in
the moderate-spin case considered in this work.  The excitation
factors described in Oshita~\cite{Oshita:2021iyn} are defined and
computed with respect to the strain. Since the NR amplitudes are
obtained from fits to the news, the first derivative of the strain, we
instead compare the measured $C_{22n}$ to $\omega_{22n} E_{22n}$.
Another important aspect in comparing the NR amplitudes to the
excitation factors of Oshita~\cite{Oshita:2021iyn} is a quantity
$t_{0}$ that defines the start of the ringdown and appears in the
equations for the excitation factors.  Since the connection between
this $t_{0}$ and the NR time coordinate is not well understood, we
simply choose a value of $t_{0}$ that produces reasonable agreement
between the amplitudes and the excitation factors.  The final freedom
in comparing the excitation factors to measured amplitudes is an
overall scale factor, since the measured amplitudes are a product of
the excitation factors and a source term, $E_{22n}T_{22n}$.  Although
the source contribution is $n$-dependent, for simplicity, we apply the
same scale factor across all $n$ by normalizing both the measured
amplitudes and the excitation factors such that $C_{220} =
\omega_{220}E_{220} = 1$.  In Fig.~\ref{fig:2420_exc_facs}, we show a
comparison between the NR amplitudes $C_{22n}$ measured in the
ringdown of SXS:BBH:2420 from Sec.~\ref{sec:MSring} and the
$\omega_{22n} E_{22n}$ associated with $\chi=0.75$, the spin of the
SXS:BBH:2420 remnant. In Fig.~\ref{fig:2420_exc_facs}, the excitation
factors are computed using a value of $t_0=2.5 M$, while the NR values
correspond to those measured at $4M$ after the peak of the $(2,2)$
component of the strain.  Overall, Fig.~\ref{fig:2420_exc_facs}, shows
that the measured NR amplitudes and the excitation factors associated
with the NR remnant are in good agreement across all $n$.

\begin{figure}[b]
  \includegraphics[width=1.0\columnwidth]{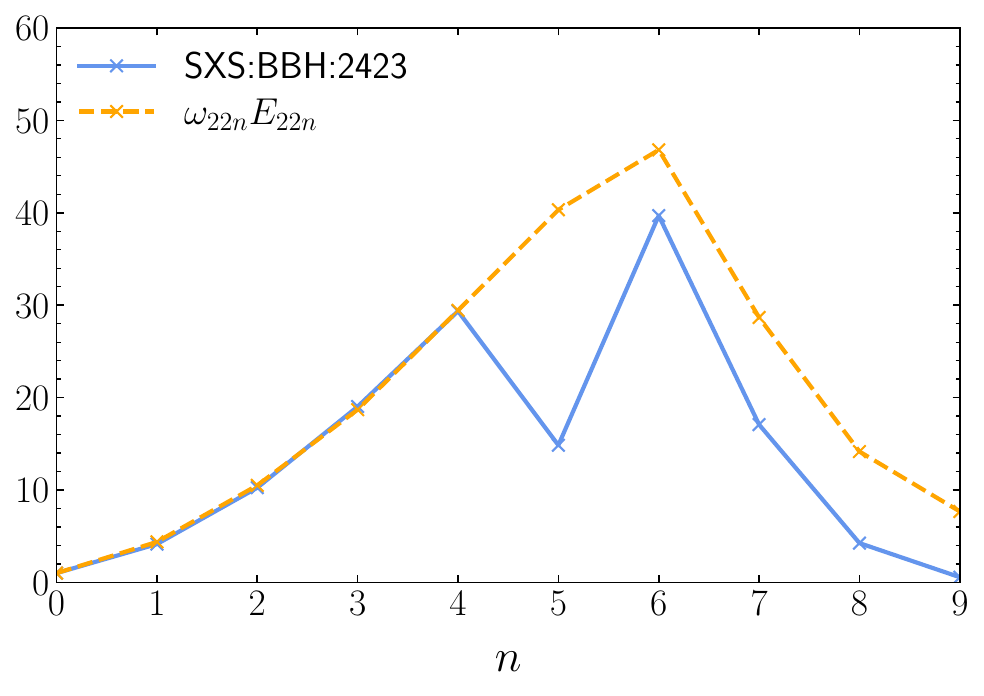}
  \caption{
    The $(2,2,n)$ NR amplitudes as measured in the
    high-spin case of SXS:BBH:2423 and the excitation
    factors $\omega_{22n} E_{22n}$ for $\chi=0.92$.
    The excitation factors are computed using
    $t_0 = 0.9$
    and both curves are normalized such that
    $C_{220} = \omega_{220}E_{220} = 1$.}
  \label{fig:2423_exc_facs}
\end{figure}

\begin{figure}[b]
  \includegraphics[width=1.0\columnwidth]{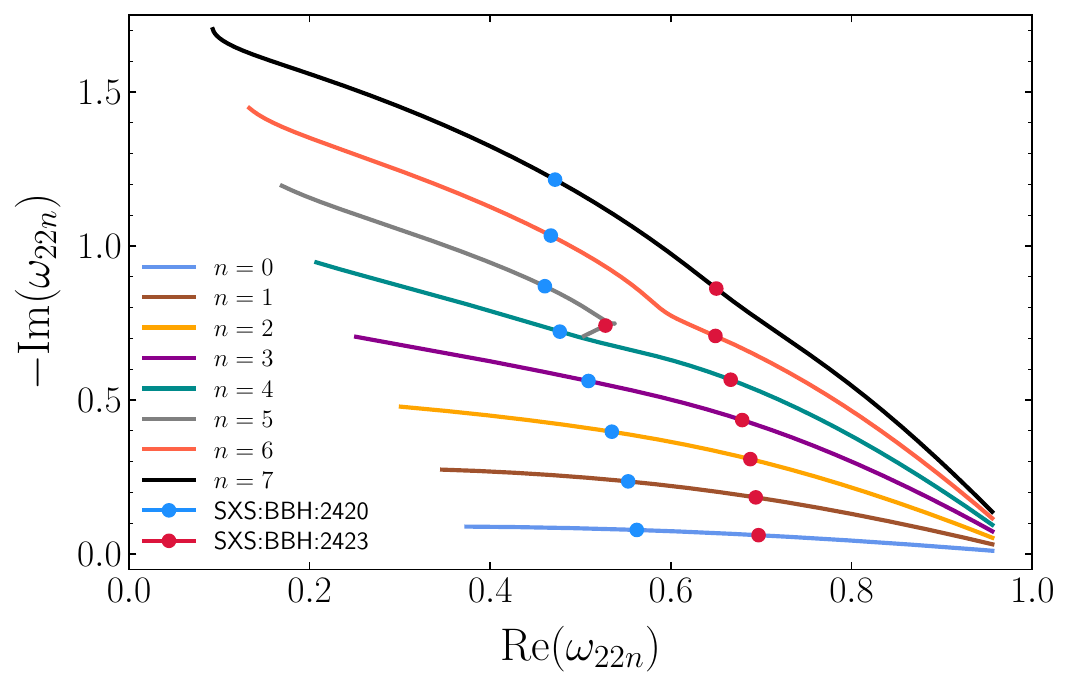}
  \caption{The $(2,2,n)$ frequencies up to $n=7$
      parameterized by $\chi$ from $0$ to $1$, with $\chi=0$
      corresponding to the leftmost point of each curve.
      Notice that all the curves, except for $n=5$, converge toward
      the same point as $\chi \rightarrow 1$.
      This divergence from the trend for $n=5$, with a
      peculiar turnaround at $\chi \sim 0.9$, is why
      the $n=5$ mode is often deemed anomalous.
      The blue dots are the QNMs associated with
      $\chi=0.76$, the remnant spin in SXS:BBH:2420,
      and the red dots are those for
      $\chi=0.92$, the remnant spin in SXS:BBH:2423.
      Notice that in the higher spin case the $n=5$
      mode becomes an outlier, remaining
      at a much lower frequency than the other $n$,
      whereas in the lower spin case the mode frequencies are
      more generally aligned.
  }
  \label{fig:22n_qnms}
\end{figure}

Returning to the case of a highly spinning remnant, we now consider
how well the excitation factors agree with the measured NR amplitudes
in the domain where $n=5$ begins to behave peculiarly.  In
Fig.~\ref{fig:2423_exc_facs}, we compare the NR amplitudes $C_{22n}$
measured in the ringdown of SXS:BBH:2423 from Sec.~\ref{sec:HSring}
and the $\omega_{22n} E_{22n}$ associated with $\chi=0.92$, the spin
of the SXS:BBH:2423 remnant. The excitation factors and measured
amplitudes are, again, both normalized. The NR amplitudes correspond
to those measured at $8M$ after the peak of the $(2,2)$ component of
the strain and a value of $t_0 = 0.9$
is used to compute the excitation
factors.  In this case, Fig.~\ref{fig:2423_exc_facs} shows excellent
agreement for $n<5$, good structural agreement for $n>5$,
while $n=5$ noticeably differs.
While the structure of the amplitudes in this case shows the
characteristic dip in the $n=5$ amplitude, the excitation factors do
not predict a dip of this magnitude until $\chi$ is a bit larger.  So
in this particular case, there is apparent disagreement between the
$n=5$ excitation factor and the NR amplitude. In the case of
SXS:BBH:2423, with a remnant spin of $\chi=0.92$, the behavior, in
particular the frequency, of the $n=5$ mode has already deviated from
the general trend that the other $n$ follow as a function of $\chi$.
As can be seen in Fig.~\ref{fig:22n_qnms},
whereas all $n \ne 5$ modes generally converge in the real part of the
frequency as $\chi \rightarrow 1$, $n=5$ becomes an outlier and
remains at a much lower frequency than the other $n$.
Given that the source terms are effectively overlap
integrals with the QNM eigenfunctions, we speculate that the anomalous
behavior of the $n=5$ mode leads to anomalous sourcing. Consequently,
the $n=5$ is sourced to a lesser extent than the other $n$.

Note that it is possible to choose to normalize the
excitation factors by an arbitrary constant scale factor,
rather than by the value for $n=0$. This normalization factor can be used
to improve the agreement with the $n=5$ dip in
Fig.~\ref{fig:2423_exc_facs} somewhat. However, the agreement
with the other values of $n$ becomes worse.

Overall, we find that the amplitudes measured in the ringdowns of NR
waveforms are generally in good agreement with the excitation factors
determined purely by $M$ and $\chi$.  Although the proposal of
comparing amplitudes to excitation factors as proof of a particular
mode measurement is interesting in theory, more work is necessary to
better understand the source factor contributions to mode amplitudes.
Given this, we leave a more rigorous investigation of the relation
between excitation factors and QNM amplitudes to future work and
instead rely on our time-dependent measurement in
Fig.~\ref{fig:2423_n5_decay} of Sec.~\ref{sec:n5} as evidence for the
presence of the anomalous $n=5$ mode.

\bibliography{references.bib}

\end{document}